\documentclass[a4paper,fleqn]{cas-dc}

\usepackage[numbers,sort&compress]{natbib}
\newcommand{\bmu}{\mbox{\boldmath $\mu$}}
\DeclareMathOperator*{\argmax}{arg\,max} 
\DeclareMathOperator*{\argmin}{arg\,min} 

\begin{document}
\let\WriteBookmarks\relax
\def\floatpagepagefraction{1}
\def\textpagefraction{.001}
\shorttitle{Optimizing sensors placement}
\shortauthors{Paluch et~al.}

\title[mode = title]{Optimizing sensors placement in complex networks for localization of~hidden signal source: A review}           

\author[1]{Robert Paluch}[bioid=1,orcid=0000-0003-4114-1806]
\credit{Conceptualization, Methodology, Software, Investigation, Writing - Original Draft}
\author[1]{\L{}ukasz G. Gajewski}[bioid=2,orcid=0000-0003-3097-0131]
\credit{Methodology,  Writing - Original Draft, Writing - Review \& Editing}
\author[1,2]{Janusz A. Ho\l{}yst}[bioid=3,orcid=0000-0003-2645-0037]
\cormark[1]
\ead{janusz.holyst@pw.edu.pl}
\credit{Supervision, Writing - Review \& Editing}
\author[3,4]{Boleslaw K. Szymanski}[bioid=4,orcid=0000-0002-0307-6743]
\credit{Resources, Supervision, Writing - Review \& Editing}

\address[1]{Warsaw University of Technology, Poland}
\address[2]{ITMO University, Sankt Petersburg, Russia}
\address[3]{Rensselaer Polytechnic Institute, USA}
\address[4]{Wroclaw University of Science and Technology, Poland}

\begin{abstract}
As the world becomes more and more interconnected, our everyday objects become part of the Internet of Things, and our lives get more and more mirrored in virtual reality, where every piece of~information, including misinformation, fake news and malware, can spread very fast practically anonymously.
To suppress such uncontrolled spread, efficient computer systems and algorithms capable to~track down such malicious information spread have to be developed.
Currently, the most effective methods for source localization are based on sensors which provide the times at which they detect the~spread.
We investigate the problem of the optimal placement of such sensors in complex networks and propose a new graph measure, called Collective Betweenness, which we compare against four other metrics.
Extensive numerical tests are performed on different types of complex networks over the wide ranges of densities of sensors and stochasticities of signal.
In these  tests, we discovered clear difference in comparative performance of the investigated optimal placement methods between real or scale-free synthetic networks versus narrow degree distribution networks.
The former have a clear region for any given method's dominance in contrast to the latter where the performance maps are less homogeneous.
We find that while choosing the best method is very network and spread dependent, there are two methods that consistently stand out.
High Variance Observers seem to do very well for spread with low stochasticity whereas Collective Betwenness, introduced in this paper, thrives when the spread is highly unpredictable.
\end{abstract}

%\includegraphics{figs/grabs.pdf}

%\item State-of-the-art and comparative study of the most popular and effective methods for sensors placement in complex networks.
%\item A new graph measure called Collective Betweenness is proposed which is designed to guide search for the most efficient placement of sensors.
%\item Extensive numerical tests have been performed on five synthetic and three real networks, for a wide range of spreading parameters (transmission variance) and densities of sensors.
%\item The results form a useful guidance for choosing the optimal method for the given network and spread process.


\maketitle

\section{Introduction}
The recent development of consumer electronics, social media and online services is changing our society with unprecedented pace.
The Internet is used no longer only for communication but also for shopping, banking, learning, entertainment, and most of all, sharing and searching for information.
The number of smart devices used in everyday life is growing, most of them connected to the Internet to~access Online Social Networks (OSN), hosted by such platforms as~Twitter, Facebook, or Instagram. The Internet became the~primary source of news, opinions and comments for many people \cite{westerman2014social}.
As always, new technologies bring not only new opportunities but also new threats.
The Internet has become a target for wide range of attacks, including propagation of~misinformation, and malware, as well as medium for fraudulent activities \cite{smith2017equi, cadwalladr2018revealed}.
Naturally, such nefarious spread is on~fundamental level similar to traditional epidemiology concerned with biological pathogen spread and therefore those two are often compared \cite{vosoughi_spread_2018, del_vicario_spreading_2016, tornberg_echo_2018, garrett_echo_2009, shao_spread_2018, shao_anatomy_2018, fraser_pandemic_2009, neumann_emergence_2009, hui2020continuing}.
While the process of spreading of biological or digital viruses has been well studied for many years \cite{PhysRevLett.86.3200,Fan2010,RevModPhys.87.925, Masood2020}, new methods for detecting and preventing malicious content in OSN are still being developed \cite{Bhattacharjee2019, Tan2019}.
The pace of this research has been accelerating recently, particularly in regard of an important problem of~locating spread source of malicious or harmful information \cite{Brockmann2013, PhysRevE.66.016128, lokhov_inferring_2014, shen_locating_2016, wang_universal_2019, Xu2019, Pinto2012, li_optimal_2020, gajewski_multiple_2019, Paluch2018, Li2019}.

Despite all these efforts, many challenges still remain. One such challenge is an optimal sensor placement to monitor the network of interest in the least expensive and unobtrusive way.
There have been plenty of recent works tackling this issue (see sec.~\ref{sec:related works} below), however, there is a lack of comprehensive, state of the art, comparative studies of this research. This motivated us to conduct such a review ourselves. In this paper, we compare performance of six algorithms (five representing the best known solutions and the sixth used as a null model, see sec.~\ref{sec:algorithms} below) used in this field in various scenarios with varying spread stochasticity, network topology and the amount of information accessible by the spread monitoring sensors.

\section{Related works}
\label{sec:related works}
Spinelli \& Celis et al. have introduced a method called High Variance Observers (sec.~\ref{sec:hvobs}) and tested it in the context of what they call a \textit{budget} and a transmission variance \cite{Spinelli2017, celis_budgeted_2015, Spinelli2019}. The budget is simply the number of sensors we are allowed to use while the transmission variance is, of~course, a measure of how non-deterministic the monitored spread is. While their results look promising, we find both the range of transmission variance and the number of networks considered not comprehensive enough.

Zhang et al. have analyzed centrality based methods and showed that none is be a clear-cut winner in terms of performance \cite{Zhang2016}. While in their work the range of sensor densities is respectable, they do not consider effects of stochasticity at all.
Based on their results, for our comparisons, we chose one of the centrality measures (sec.~\ref{sec:bc}) that seems slightly better than the rest of~them.

In contrast to above work, we consider both - sensor density and~transmission variance - in a wide range of values and on top of~that we use eight different networks as our testing environments.

There have also been some works on {\it online} sensors selection to be able to take into account the spread evolving dynamics \cite{Spinelli:2017:BSO:3038912.3052584, Spinelli2019, Zhang2016/05}. However, we consider such approach to be beyond the scope of our study because it introduces new issues and challenges.

On~similar note there has been some impressive work by Zejnilovic et al. \cite{zejnilovic_sequential_2015, zejnilovic_selecting_2015, zejnilovic_network_2013}, Wang \cite{wang_universal_2019}, Fang et al. \cite{Fang2018}, Li et~al. \cite{Li2019} and Shi et al. \cite{Shi2019} where new localization methods are introduced and several methods for sensor placement are studied.
Nevertheless, the localization scheme and, most importantly, certain assumptions are different than the ones we are using in this paper and therefore we will omit those as~well.

\section{Basics}

\subsection{Spreading model}

We use Susceptible-Infected model~\cite{bailey1975mathematical} to simulate propagation over the complex network.
SI model has only one parameter, \emph{infection rate} $\beta$, which is a probability per time step that infected node will transmit the infection to the uninfected neighbor.
The infected nodes try to infect their neighbors in every time step.
The distribution of the number of time steps needed to transmit the infection is geometric, with $\mu=1/\beta$ and $\sigma=\sqrt{1-\beta}/\beta$.
We define \emph{transmission variance} $\xi$ according to Spinelli et al. \cite{Spinelli2017}, as the ratio between the standard deviation and the mean of the number of times steps needed to transmit the infection $\xi=\sigma/\mu=\sqrt{1-\beta}$.
Please note that $\xi$ is nothing else than inverse of propagation ratio $\lambda$ \cite{Pinto2012,Paluch2018}.

\subsection{Source localization algorithm}
For the localization of spread source, Pinto-\allowbreak Thiran-\allowbreak Vetterli \cite{Pinto2012} algorithm is used in restricted form (PTVA-LI~\cite{Paluch2018}).

Let us have a network $G = (V, E)$, with $V, E$ being its known sets of vertices (nodes) and edges (links) respectively, defining our system. In this system, we place our set of sensors $S \subset V$, that is the nodes that report at what time they got infected, and an unknown origin of infection $o^\ast$. Normally (i.e., in PTVA), we would also register from whom given sensor received the virus, however, this is where the \textit{restricted form} part comes in - we use \textit{limited information} version, namely PTVA-LI and only the times of infection are given by the sensors.

We assume that the inception time of the virus $t_0$ is unknown, and only the mean and variance of the transmission time $\mu, \sigma$ per link are known (but not exact propagation times).

The goal, of course, is to locate $o^\ast$

From infection times reported by sensors we construct an observed delay vector $\mathbf{d}$:
\begin{equation}
    \mathbf{d} = (t_2 - t_1, t_3 - t_1, \dots, t_{b} - t_1)^T
\end{equation}
where $b$ is the number of sensors (budget), $t_i$ is an infection time of sensor $s_i \in S$, and $t_1$ is the infection time of a \textit{reference sensor} that is needed here since the $t_0$ is unknown.

For each node in the system $v \in V$ we compute a deterministic delay vector \bmu:
\begin{equation}
\label{eq:det_delay}
    \bmu_v = \mu \big( |P(v, s_2)| - |P(v, s_1)|, \dots,
    |P(v, s_{b})| - |P(v, o_1)| \big)^T
\end{equation}
where $|P(v, s_i)|$ is number of edges on a shortest path connecting nodes $v, s_i$. We also compute the covariance matrix $\mathbf{\Lambda}_v$, each element $i,j$ of which is given by:
\begin{equation}
    \mathbf{\Lambda}_{i,j} = \sigma^2 \times \begin{cases} |P(s_i, s_1)| & i = j,
    \\  |P(s_i, s_1) \cap P(s_j, s_1)| & i \neq j
     \end{cases}
\end{equation}
Finally we compute a \textit{score} for each node $v$ and use maximum likelihood rule to determine the most probable origin of the epidemic $\hat{o}$:
\begin{equation}
\label{eq:argmax}
    \hat{o} = \argmax_{v \in V} \bmu^T_v \mathbf{\Lambda}^{-1}_v (\mathbf{d} - 0.5\bmu_v)
\end{equation}
When $\hat{o} = o^\ast$ we count it is as a success. See evaluation metrics below for details.

Since in general $G$ can be any graph and PTVA is optimal on trees, one must construct a BFS (breadth first search) tree on each node $v \in V$ and apply the above described procedure (eq.~(\ref{eq:det_delay})-(\ref{eq:argmax})) on each tree respectively.

\section{Algorithms}
\label{sec:algorithms}
We compare the following six methods of efficient sensors' selection.

\subsection{Betweenness Centrality (BC)}
\label{sec:bc}
This popular and simple heuristic takes the nodes with largest betweenness centrality, which is computed independently for every node $v \in V$ as
\begin{equation}
    %\text{BC}(v) = \sum_{i,j \in V, i\neq j \neq v} \frac{\sigma_{ij}(v)}{\sigma_{ij}},
    S_{\text{BC}} = \argmax_{S} \sum_{v \in S} \sum_{\substack{i,j \in V \\ i\neq j \neq v}} \sigma_{ij}^{(v)}/\sigma_{ij},
\end{equation}
where $\sigma_{ij}^{(v)}$ is the number of shortest paths between $i$ and $j$ which contain $v$ and $\sigma_{ij}$ is the total number of shortest paths between them.

\subsection{High Coverage Rate (Coverage)}
The algorithm, proposed by Zhang et al., selects a set of the sensors $S_\text{Coverage}$ which has the maximum number of unique neighbors \cite{Zhang2016}.
The nodes which have an sensor as a neighbor are \textit{covered}, and the fraction of covered nodes in the network is called \textit{coverage rate}.
This method saturates, since usually the density of sensors which gives coverage rate equal one is significantly smaller than one.

To avoid the problem of saturation, additional stages are added to the algorithm in the following manner.
The first stage is identical as originally proposed by Zhang et al.~\cite{Zhang2016}.
The sensors are chosen greedily, one by one, and each new sensor increases the coverage rate until it reaches unity.
When the coverage rate became one, the next stage starts.
In the second stage, the algorithm  selects sensors which maximize the number of nodes which have two sensors as the neighbors (\textit{double-covered} nodes).
In the third stage algorithm maximizes the number of \textit{triple-covered} nodes and so on until the desirable density of sensors is reached.

\subsection{K-Median (K-Median)}
K-Median placement was proposed by Berry et al. for efficient detectability of a flow in municipal water networks \cite{Berry2006}.
This method minimizes the sum of distances between the nodes and the closest sensors.
If $V$ is the set of all nodes, $d(i,j)$ is the length of the shortest path between nodes $i$ and $j$, $S_{\text{K-Median}}$ set of sensors is
\begin{equation}
    S_{\text{K-Median}} = \argmin_{S}{\sum_{i \in V} \min_{o \in S}{d(i,o)}}
\end{equation}

\subsection{High Variance Observers (HV-Obs)}
\label{sec:hvobs}
The algorithm introduced by Spinelli et al. is based~on path covering strategy \cite{Spinelli2017}.
This method looks for a set of sensors $S_{\text{HV-Obs}(L)}$ that maximizes the cardinality of $P_L(S)$, which is the set of nodes that lie on a shortest paths of length at most $L$ between any two sensors in the set $S$.

High Variance Observers is designed for small density of sensors, since $|P_L(S)|$ increases quickly with the number of sensors and can easily reach the number of all nodes in the network.
Here, to solve this problem, we extend the algorithm in the similar way as in the case of High Coverage Rate method.
A node which lies on an exactly one shortest path of length at most $L$ between any two sensors in the set $S$ is called \textit{single-path-covered}.
A \textit{double-path-covered} node lies on two shortest paths, \textit{triple-path-covered} lies on three shortest paths and so on.
In the first stage, the algorithm selects greedily the sensors which maximize the number of \textit{single-path-covered} nodes until all nodes are \textit{single-path-covered}.
Then, the second stage starts, in which the number of \textit{double-path-covered} is maximized and so on until the desirable density of sensors is reached.

\subsection{Collective Betweenness (CB)}
We propose a novel method which maximizes a new measure called Collective Betweenness:
\begin{equation}
    %\text{CB}(K) = \sum_{i,j \in V, i\neq j \neq v} \frac{\sigma_{ij}(K)}{\sigma_{ij}},
    S_{\text{CB}} = \argmax_{S} \sum_{\substack{i,j \in V \\ i,j \notin S}} \sigma_{ij}^{(S)}/\sigma_{ij},
\end{equation}
where $S$ denotes a set of sensors, $\sigma_{ij}^{(S)}$ is the number of shortest paths between $i$ and $j$ which pass through any node belonging to $S$ and $\sigma_{ij}$ is the total number of shortest paths between them.
In this approach each shortest path is counted only once (even if it passes through many nodes belonging to $S$), therefore value of CB for the set $S$ is different than simple sum of Betweenness Centralities of all nodes in $S$.

\subsection{Random}
This is a benchmark for the rest of methods (a baseline). This method selects sensors randomly.

\subsection{Complexity}
\begin{figure}
    \centering
    \includegraphics[width=0.99\columnwidth]{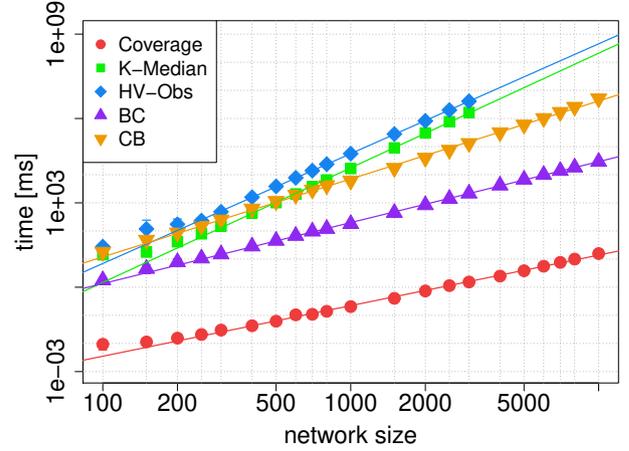}
    \caption{Average time needed for determining the sensors set (density of sensors $\rho=0.05$) versus the size of Erd\H{o}s-R\'{e}nyi network with average degree $\langle k \rangle = 8$.
    Error bars representing confidence interval of level $0.95$ are smaller than symbols in most cases.
    Solid lines are linear models $\ln(\text{time}) = \gamma \ln(\text{size}) + \text{const}$.
    The values of $\gamma$ are presented in Table \ref{tab:complexity}. HV-Obs and K-Median were parallelized using 16 threads.}
    \label{fig:complexity}
\end{figure}

Computational complexity is an important criterion for the usability of the algorithms.
As seen in Table~\ref{tab:complexity}, time complexity of methods studied in this article varies from  $O(nm)$ to $O(b^2n^2)$, where $n=|V|$ is the number of nodes, $m=|E|$ is the number of links and $b=|S|$ is the number of sensors (budget).
The third column contains the results of~numerical experiment in which the average execution time for each method was measured as a function of network size (see with Fig. \ref{fig:complexity}).
The experiment was conducted for Erd\H{o}s-R\'{e}nyi network with constant density of sensors $\rho=0.05$ and average degree $\langle k \rangle = 8$.
High Coverage Rate has the lowest computational complexity among the tested methods.
The authors state that the complexity of this algorithm can be reduced to $O(n+m)$ \cite{Zhang2016}, but our version, which is resistant to~the~saturation effect, has complexity $O(bn)$.
On the opposite extreme, there are High Variance Observers and K-Median.
In~case of these algorithms, the maximum network size during the experiment shown in Fig. \ref{fig:complexity} is $3000$ due to~very long time of computations.
Although both methods are well parallelizable, the maximum possible speed-up is equal to the~number of nodes in the network (and requires this number of threads), which can be still insufficient for applying these algorithms to large networks.

\begin{table*}
    \centering
    \caption{Computational complexity of studied algorithms. Here $n=|V|$ is the number of nodes, $m=|E|$ is the number of links, $b=|S|$ is the number of sensors (budget), and $\gamma$ is an experimentally obtained scaling exponent from the linear model best fitting to results of numerical simulations (see Fig. \ref{fig:complexity}).}
    \begin{tabular}{llll}
        \hline
        Algorithm & Complexity & $\gamma$ & Parallelizable?\\
        \hline
        High Coverage Rate \cite{Zhang2016} & $O(bn)$ & 1.80(2) & No\\
        Betweenness Centrality \cite{Brandes2001} & $O(nm)$ & 2.17(1) & Yes\\
        Collective Betweenness & $O(nm+bn\log^2n)$ & 2.77(3) & Partially\\
        High Variance Observers \cite{Spinelli2017} & $O(b^2n^2)$ & 3.91(1) & Yes\\
        K-Median \cite{Kariv1979} & $O(b^2n^2)$ & 4.08(3) & Yes\\
        \hline
    \end{tabular}
    \label{tab:complexity}
\end{table*}

\subsection{Similarity between sets of sensors}
\begin{figure*}
    \centering
    \includegraphics[width=0.85\textwidth]{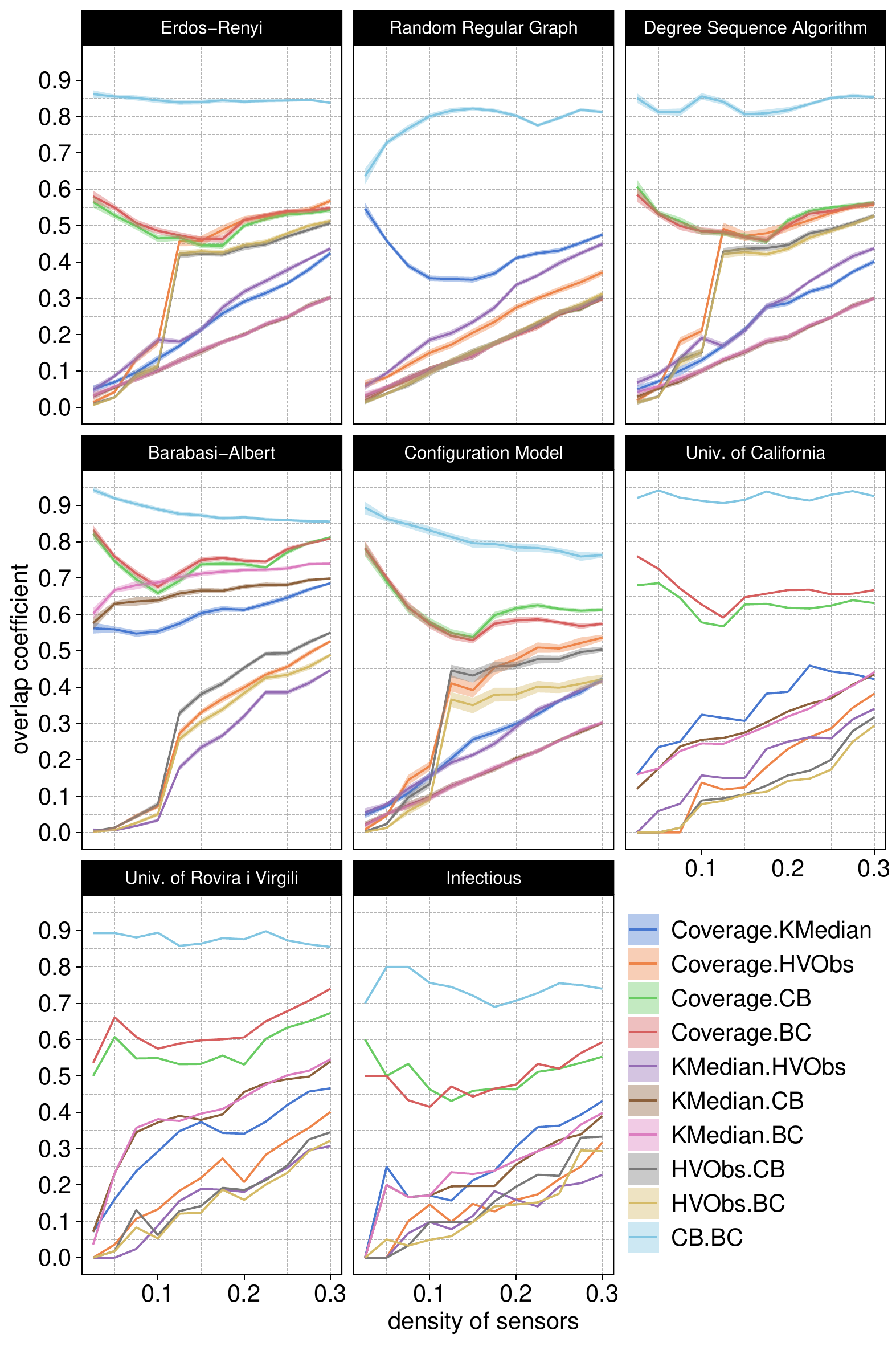}
    \caption{The overlap coefficient as a function of density of sensors for all networks used in this study.
    All panels have the same horizontal axis.
    The results for synthetic networks are averaged over 50-150 realizations. Error bands represent the confidence intervals at the level $0.95$.}
    \label{fig:overlap}
\end{figure*}

The algorithms for sensor placement are very different (except for BC and CB), yet, the sets of nodes which they generate often are very similar to each other.
We use an overlap coefficient \cite{Vijaymeena2016} as similarity measure:
\begin{equation}
    \text{overlap}(X,Y) = \frac{|X \cap Y|}{\min(|X|,|Y|)} = \frac{|X \cap Y|}{s},
\end{equation}
where $s=|X|=|Y|$ is the number of sensors.
Figure \ref{fig:overlap} presents the overlap coefficient as a function of the density of sensors for all networks used in this study.
The overlap coefficients between random sensors and other sets of sensors are not shown because their values are known and equal to the density of sensors $\rho$.

As expected, the highest overlap occurs for Betweenness Centrality and Collective Betweenness.
It oscillates between 0.7 for the Infectious network and 0.95 for the University of California network.
The most unique set of sensors for every network is the set given by High Variance Observers method.
Any overlap with this method is always below 0.55.

\section{Evaluation metrics}

Two efficiency measures are used for evaluating the quality of origin detection: the average precision and the Credible Set Size at $0.95$ confidence level.
The precision for a single test is defined as the ratio between the number of correctly located sources (i.e., true positives, which here equals either zero or one) and the number of sources found by the method (i.e., true positives plus false positives, which here is at least one).
The tests are repeated multiple times for different origins and many graph realizations (for~synthetic networks) and then the obtained values of precision are averaged.
The Credible Set Size at the confidence level of $\alpha$ ($\alpha\text{-CSS}$) is a novel metric introduced by us here. It is the size of the smallest set of nodes containing the true source with probability $\alpha$.
In other words this metric describes how many nodes with the highest \textit{score} should be labeled as origin to have probability $\alpha$ that the true origin is among these nodes.
Probability here is understood as frequency and computed as a hit rate (recall) from many realizations of signal propagation and source location.

\section{Results}

We evaluate the algorithms for sensor placement on several synthetic networks, such as Erd\H{o}s-R\'{e}nyi model \cite{Erdos1960}, Random Regular Graph, Degree Sequence Algorithm \cite{Heath2011} with Poisson distribution, Barab\'{a}si-Albert model \cite{Albert2002}, Configuration Model with power-law degree distribution and three real networks (one network of human face-to-face contacts and two networks of Internet communications between academics).
We study how the transmission variance $\xi$ influences the quality of source detection for various densities of sensors $\rho$.
We investigate all networks as unweighted to limit the space of possible models’ parameters.
The heterogeneity of links is partially incorporated in the stochastic character of spreading process.

Figures \ref{fig:er_results}-\ref{fig:infect_results} show the results of numerical simulations in the most straightforward non-aggregated form.
In~each figure, the top two rows of charts present average precision, and the bottom two rows present Credible Set Size at the confidence level of $0.95$ as a function of transmission variance $\xi$ for six values of sensors' density (5\%-30\%).
As evident, the transmission variance is a major factor in the quality of source detection.
For the low transmission variance, the precision provided by any reasonable algorithm for sensor placement is very high.
In~the opposite case, for the very high transmission variance, the quality of all methods is rather poor.
The~middle range of the variance transmission is characterized by the largest differences in precision and 0.95-CSS between the tested algorithms.
The~limit of this range depends on the density of~sensors -- it shifts towards the higher values of transmission variance with the higher density of sensors.
This behavior can be observed for all types of synthetic and some real networks.

Summary Diagrams presented in Figures \ref{fig:er_diagrams}-\ref{fig:infect_diagrams} visualise extensive amount of complex results in an~innovative way.
In~each figure, the top diagram refers to average precision, while the bottom diagram relates to 0.95-CSS. 
Each diagram consists of $96$ tiles ($16$ values of $\xi$ and $6$ values of $\rho$).
The color of the background in each tile indicates which algorithm provides the highest average precision (top) or the smallest 0.95-CSS (bottom) for the given pair ($\xi,\rho$).
Moreover, inside each tile there are five colored bars, ordered from highest to lowest, that illustrate the ranking of the methods.
The first bar relates to the best algorithm in the given tile, the second refers to second best and so on.
The sixth algorithm in given tile is not shown, since it height would be zero.
This is because the height of each bar shows the difference in average precision (or 0.95-CSS) between the method to which this bar refers and the worst (sixth) method in given tile.
Then, the heights of bars from all tiles are scaled relative to the height of the highest bar among them, with a minimum height to recognize the color.
This calibration allows to compare bars from various region of parameter space at cost of comparison of bars within one tile.
Summary Diagrams give quick insight into results and show in legible way which algorithms lead in which regions of the parameter space.

To better understand and interpret these complex results, we partition the parameter space into nine regions of similar size and shared boundaries which correspond to low, medium and high transmission variance and density of sensors.
Tables \ref{tab:er}-\ref{tab:infect} present average values for precision (top numbers in cells, in percentages) and Credible Set Size (bottom numbers in cells) in these regions.
In each table, the first three numerical columns from the left refer to low, three in the middle to medium, and the last three to high transmission variance $\xi$.
Similarly, columns $\{1,4,7\}$ correspond to high, $\{2,5,8\}$ to middle, and $\{3,6,9\}$ to low density of sensors $\rho$.
This arrangement of columns causes the precision decreases from left to right side of the table.
The best results in each region are printed in bold. The uncertainty of precision is given by the confidence interval at the level $0.95$.

\subsection{Tests on synthetic networks}

For each model of network and each value of $\rho$ we execute the following script $100$ times:
    \begin{enumerate}
        \item Generate a new graph.
        \item Find sets of sensors according to six different selection strategies (Random, Coverage, K-Median, HV-Obs, BC and CB).
        \item For each value of $\xi$ repeat $100$ times:
            \begin{enumerate}
                \item Simulate spread from random source using SI model.
                \item Locate the source six times using different sets of sensors. 
            \end{enumerate}
    \end{enumerate}
As a result, the values of average precision and the average Credible Set Size at the confidence level $0.95$ for each point ($\rho,\xi$) are computed from $10^4$ attempts to locate the source.
Table \ref{tab:synthetic} presents the characteristics of studied networks.

\begin{table*}
    \centering
    \caption{The number of nodes in each network $n=1000$. The values of the average degree $\langle k \rangle$, the average path length (APL) and the global clustering coefficient are averages of $100$ realizations of graphs.}
    \begin{tabular}{lllll}
        \hline
        Graph type & Degree distribution & $\langle k \rangle$ & APL & Global clustering\\
        \hline
        Erd\H{o}s-R\'{e}nyi & Binomial & $8.00$ & $3.55$ & $0.008$ \\
        Random Regular Graph & $k=\text{const}$ & $8.00$ & $3.60$ & $0.006$\\
        Degree Sequence Algorithm & Poisson & $7.67$ & $3.76$ & $0.132$\\
        Barab\'{a}si-Albert & $P(k) \sim k^{-3}$ & $7.98$ & $3.17$ & $0.026$ \\
        Configuration Model & $P(k) \sim k^{-3}$ & $7.02$ & $3.36$ & $0.021$ \\
        \hline
    \end{tabular}
    \label{tab:synthetic}
\end{table*}

\subsubsection{Erd\H{o}s-R\'{e}nyi Graph (ER)}
This random graph is constructed by connecting every pair of nodes with probability $p$.
The resulting network has binomial degree distribution with the average degree $\langle k \rangle = pn$.
The average path length (APL) scales linearly with $\ln{n}$ (it is a property of so-called \textit{small-world networks}) and it is much smaller than the number of edges.
The global clustering coefficient is $p$ since the probability that two connected nodes have a common neighbor is uniform for all nodes and equal to $p$.
According to Table \ref{tab:er}, Coverage and K-Median are the best approaches for sensor placement in ER graph for the moderate transmission variance rage $\xi \in \langle 0.5;0.8 \rangle$ and the density of sensors below or equal to 20\%.
For larger budgets $\rho \in \langle 20\%;30\% \rangle$ Collective Betweenness delivers the highest quality of source detection in that range of transmission variance.
Also CB is the very effective when spread is highly stochastic $\xi \in \langle 0.8;0.95 \rangle$ and the density of sensors higher or equal to 10\%.
For the low transmission variance range all methods give the same values for the Credible Set Sizes with the confidence level $0.95$, but K-Median has the highest average precision.

\subsubsection{Random Regular Graph (RRG)}
Nodes in RRG are connected at random, but with the constraint that each node has the same degree.
RRG has a slightly lower global clustering coefficient and a higher average path length than ER graph with the same number of nodes and edges.
Table \ref{tab:rrg} shows that K-Median and Coverage provides the highest quality of source detection, but for the most stochastic processes and largest budgets, Collective Betweenness gives the smallest Credible Set Sizes at the confidence level $0.95$ and it is the second best in the average precision.
In this region (the highest transmission variance and density of sensors), difference between the efficiency of Collective Betweenness and Betweenness Centrality is the largest among all studied networks and all regions.
In fact, for $\xi=0.95$ and $\rho \in \langle 25\%,30\% \rangle$ CB provides the highest average precision, while BC is the second worse in that region (see Fig.\ref{fig:rrg_diagrams}).

\subsubsection{Degree Sequence Algorithm (DSA)}
This model is a random synthetic network with the high clustering coefficient.
The algorithm takes a sequence of nodes' degrees as an input and global clustering coefficient as parameter, and returns the list of links forming the graph.
The degrees of nodes are selected from the Poisson distribution for easier comparison with ER graph.
The results presented in Table \ref{tab:pois} indicate a significant advantage of Collective Betweenness over the other methods for the transmission variance $\xi \geqslant 0.5$ and the density of sensors $\rho \geqslant 10\%$.
For the smaller budgets $\rho \leqslant 10\%$, Coverage and Betweenness Centrality provide higher precision than CB.
In the case of low transmission variance $\xi \leqslant 0.5$, HV-Obs gives the highest precision for $\rho \geqslant 20\%$.
In contrast to the results for ER and RRG, here K-Median performs poorly, with exception of cases with the lowest values of transmission variance and the largest numbers of sensors.

\subsubsection{Barab\'{a}si-Albert Model (BA)}
This algorithm uses the preferential attachment rule to generate a scale-free network, which has a smaller average path length and a higher clustering coefficient than Erd\H{o}s-R\'{e}nyi graph (but sill much smaller than for real social networks).
Figure \ref{fig:ba_diagrams} reveals fragmentation of parameter space into three distinct regions.
The first region includes all the densities of sensors with the transmission variance $\xi < 0.5$.
In this region, HV-Obs provides the highest quality of source localization (the lower density of sensors is, the bigger is the advantage of HV-Obs over competitors).
The second region, in which CB is the most effective method, includes all the densities of sensors with the transmission variance $\xi > 0.5$, except of the corner with the highest transmission variance and the lowest density of sensors, where BC is the best method.

\subsubsection{Configuration Model (CM)}
This model generates a random graph from a  given degree sequence.
In our studies, we use the power law degree distribution for easier comparison with the Barab\'{a}si-Albert model.
Although the results shown in Table \ref{tab:sfn} are consistent with the results for BA network, Fig. \ref{fig:sfn_diagrams} reveals a small subregion (with the medium transmission variance $\sigma=0.55$ and the high density of sensors $\rho \geqslant 15\%$) where the leading method is Coverage.

\subsubsection{Summary of tests on synthetic networks}
The results in tables \ref{tab:er}-\ref{tab:sfn} show that for the low transmission variance $\xi \in \langle 0.2;0.5 \rangle$, the profit from choosing the particular set of sensors is very small when the density of sensors exceed 10\%.
In the case of lower density of sensors, the highest quality of source detection is provided by HV-Obs (for the Degree Sequence Algorithm, Configuration Model, Barab\'{a}si-Albert model) and K-Median (for the Erd\H{o}s-R\'{e}nyi graph, Random Regular Graph).
On the opposite side, for the high transmission variance $\xi \in \langle 0.8;0.95 \rangle$, the gain from using specific sensors is much higher, but the choice of algorithm depends on the network type.
For Erd\H{o}s-R\'{e}nyi the best are Coverage (when $\rho \leqslant 10\%$) and Collective Betweenness (when $\rho>10\%$). 
Coverage is also the best choice for Random Regular Graph. 
For the Degree Sequence Algorithm, Barab\'{a}si-Albert and Configuration Model, the highest precision and the smallest Credible Set Size at the confidence level $0.95$ are provided by the Collective Betweenness.
Also in the middle range of transmission variance $\xi \in \langle 0.5;0.8 \rangle$ 
The Collective Betweenness is the best algorithm for these graphs (except DSA with low density of sensors), but for ER and RRG it gives way to Coverage and K-Median.

\subsection{Tests on real networks}
The real networks used in the study come from the Koblenz Network Collection~\cite{Kunegis2013} (KONTECT).
Table \ref{tab:real} presents the characteristics of these networks.
For each real network and each pair of $(\rho,\xi)$ we simulate the spread from a random source and locate the source using different sets of sensors $10^4$ times.

\begin{table*}
    \centering
    \caption{Basic properties of the real networks used in tests.}
    \begin{tabular}{lllllll}
        \hline
        Network & |V| & $\langle k \rangle$ & $k_{max}$ & APL & Diameter & Global clustering \\
        \hline
        Univ. of California & 1020 & 12.2 & 110 & 3.0 & 5 & 0.046\\
        Univ. of Rovira i Virgili & 1133 & 9.6 & 71 & 3.6 & 8 & 0.166\\
        Infectious & 410 & 13.5 & 50 & 3.6 & 9 & 0.436\\
        \hline
    \end{tabular}
    \label{tab:real}
\end{table*}

\subsubsection{University of California}
This network contains information about the message exchanges between the users of an online community of students from the University of California, Irvine \cite{Opsahl2009}
A node represents a user and the directed multiple edges represent messages.
We transform the network to undirected one in the same way as Spinelli et al.~\cite{Spinelli2017}, by aggregating all edges between a given pair of nodes and leaving only the connections with at least one edge in both directions.
Then we remove iteratively all nodes with less than two connections until the minimum node degree in the network is two.
The results of numerical tests contained in Table \ref{tab:uc} are very similar to the results for Barab\'{a}si-Albert model (Table \ref{tab:ba}).
The most effective method for the low transmission variance $\xi$ is HV-Obs, while for the medium and high $\xi$, the best quality of source detection is provided by Betweenness Centrality and Collective Betweenness.

\subsubsection{University of Rovira i Virgili}
This is the email communication network at the University Rovira i Virgili in Tarragona in the south of Catalonia in Spain~\cite{Guimera2003}.
An undirected link between two nodes (users) is created when at least one email was sent from one user to another.
The results obtained for this network are in line with the results for synthetic scale-free networks and the University of California network.
The main difference is observed for the low transmission variance $\xi \in \langle 0.2;0.5 \rangle$ and $\rho \geqslant 10\%$, where all sets of sensors perform worse than random ones (see Table \ref{tab:rovira}).

\subsubsection{Infectious}
This network contains human face-to-face interactions during the exhibition INFECTIOUS: STAY AWAY in 2009 at the Science Gallery in Dublin~\cite{Isella2011}.
Edges represent contacts which lasted for at least 20 seconds.
Only the data from the day with the most interactions was used.
The network is characterized by highest average degree and clustering coefficient among all tested graphs.
Table \ref{tab:infect} shows that the low transmission variance region is again dominated by HV-Obs, in particular for $\rho \leqslant 10\%$.
The best option for the high transmission variance is Betweenness Centrality, and Collective Betweenness is the first choice for medium range of $\xi$, except of a region with the low density of sensors $\rho \leqslant 10\%$, where Coverage performs better.

\subsubsection{Summary of tests on real networks}
In our tests, locating the source of information is much more challenging on real networks than on  artificial ones due to high clustering coefficient of such networks.
On the other hand, employing the right strategy of sensor placement can be more fruitful in case of the real systems than for the synthetic ones.
The common denominator for all the real networks studied in this paper is the advantage of betweenness-based algorithms over the rest of methods for the medium and high transmission variance.
In the case of low transmission variance, best strategy for sensor placement are HV-Obs and Random (the latter only for the University of Rovira i Virgili network).
For the Infectious network also Coverage performs well for medium transmission variance (in particular for $\xi=0.55$).

\section{Discussion}
In this article we review the methods of sensor placement for the source localization in complex networks, both real and synthetic, using a well established propagation model - Susceptible-Infected - and the Pinto-Thiran-Vetterli localization algorithm.
We have selected four vastly acknowledged methods for sensor placement - Betweenness Centrality (BC), High Coverage Rate (Coverage), K-Median, High Variance Observers (HV-Obs) - and also have introduced our own method called Collective Betweenness (CB). We compare all of these methods with each other
and with a baseline method - random selection.
As our main evaluation metrics, we use an average precision of identifying the actual source of the spread and introduce a new  metric called Credible Set Size that we believe to be more useful in possible real world scenarios as it conveys a very practical notion of ``how many nodes do I need to check to say the source is one of them with credibility $\alpha$?".
The study was conducted over a large range of values for the two main parameters affecting the propagation and source identification - the transmission variance $\xi$ and the density of sensors $\rho$.

\begin{figure*}
    \centering
    \includegraphics[width=0.96\textwidth]{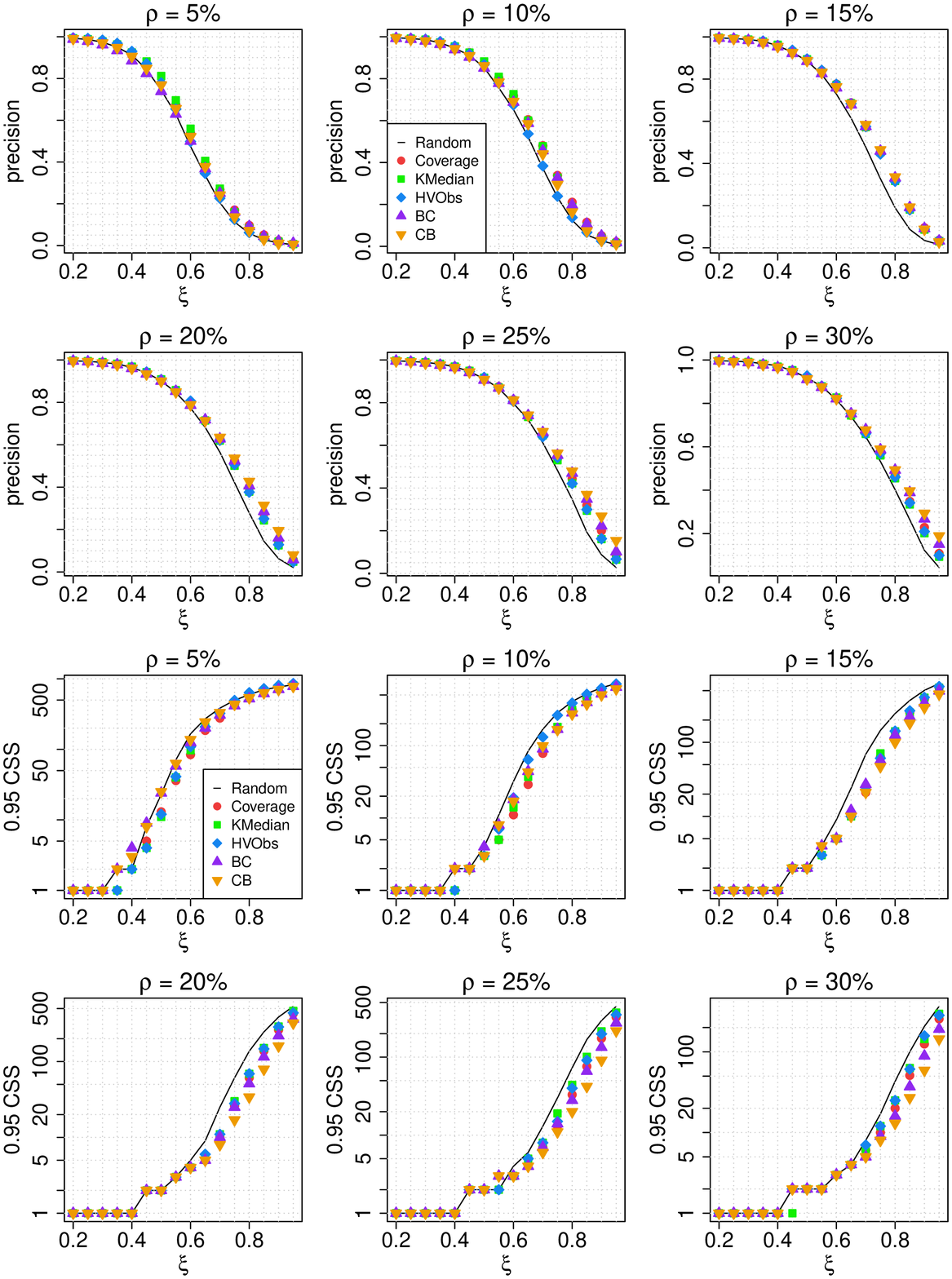}
    \caption{Comparison of source localization quality for five sensor placement strategies in case of \textbf{Erd\H{o}s-R\'{e}nyi graph} ($n=1000$, $\langle k \rangle = 8$).
             A black solid line denotes the results for randomly placed sensors.
             The confidence intervals at the level $0.95$ are smaller than symbols.}
    \label{fig:er_results}
\end{figure*}

\begin{figure*}
    \centering
    \includegraphics[width=0.96\textwidth]{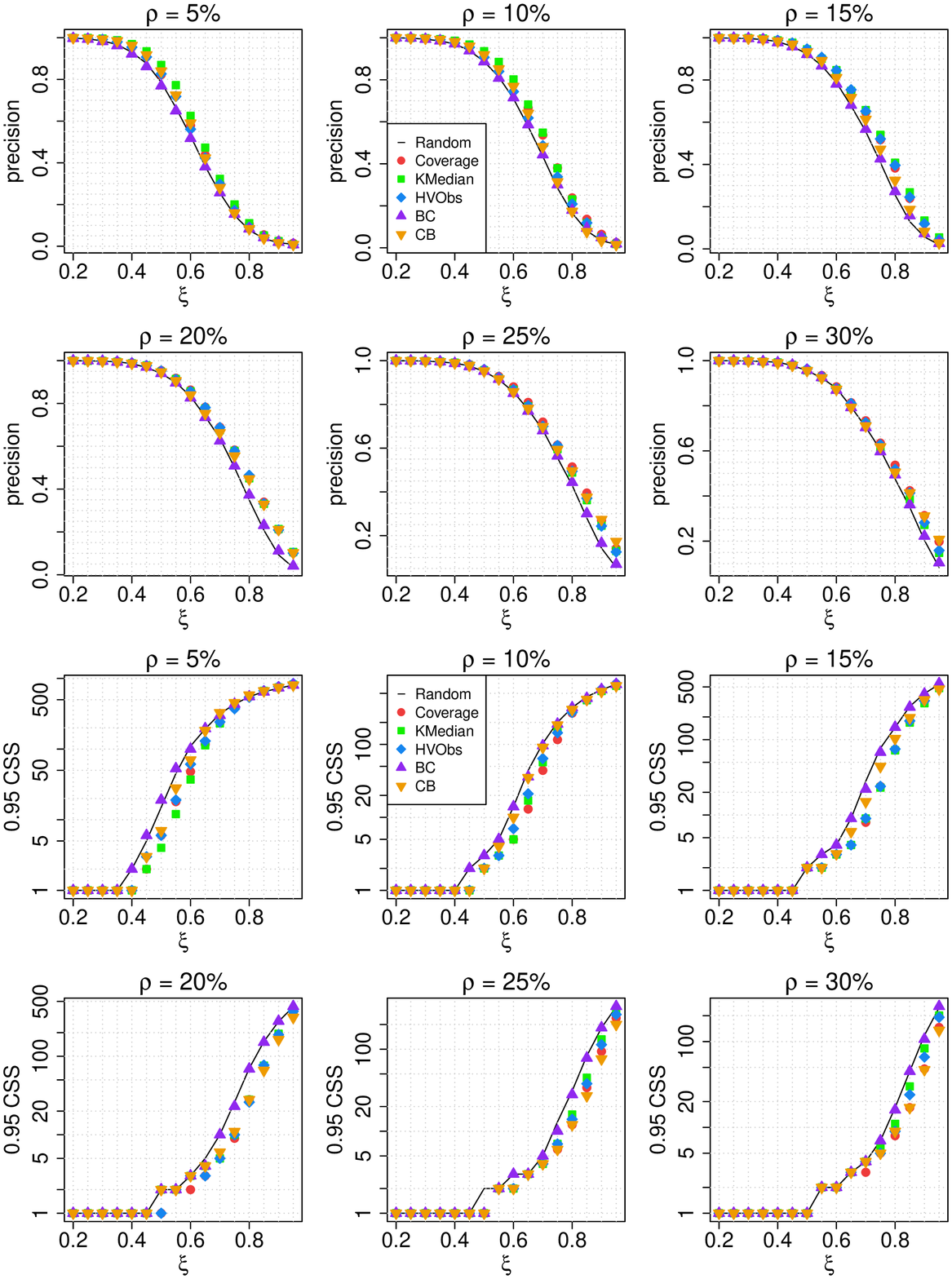}
    \caption{Comparison of source localization quality for five sensor placement strategies in case of \textbf{Random Regular Graph} ($n=1000$, $\langle k \rangle = 8$).
             A black solid line denotes the results for randomly placed sensors.
             The confidence intervals at the level $0.95$ are smaller than symbols.}
    \label{fig:rrg_results}
\end{figure*}

\begin{figure*}
    \centering
    \includegraphics[width=0.96\textwidth]{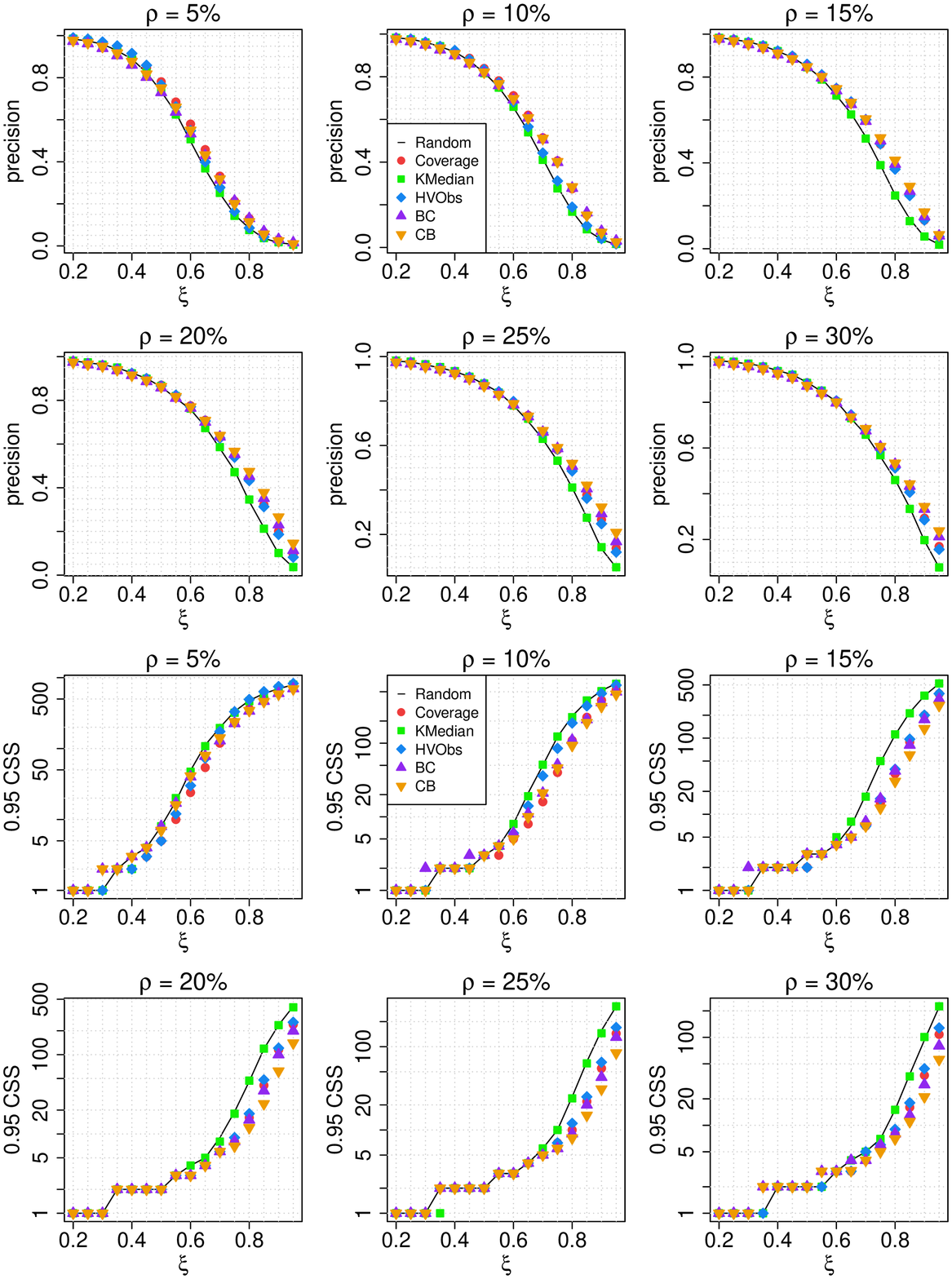}
    \caption{Comparison of source localization quality for five sensor placement strategies in case of \textbf{Degree Sequence Algorithm} ($n=1000$, $\langle k \rangle = 7.67$).
             A black solid line denotes the results for randomly placed sensors.
             The confidence intervals at the level $0.95$ are smaller than symbols.}
    \label{fig:pois_results}
\end{figure*}

\begin{figure*}
    \centering
    \includegraphics[width=0.96\textwidth]{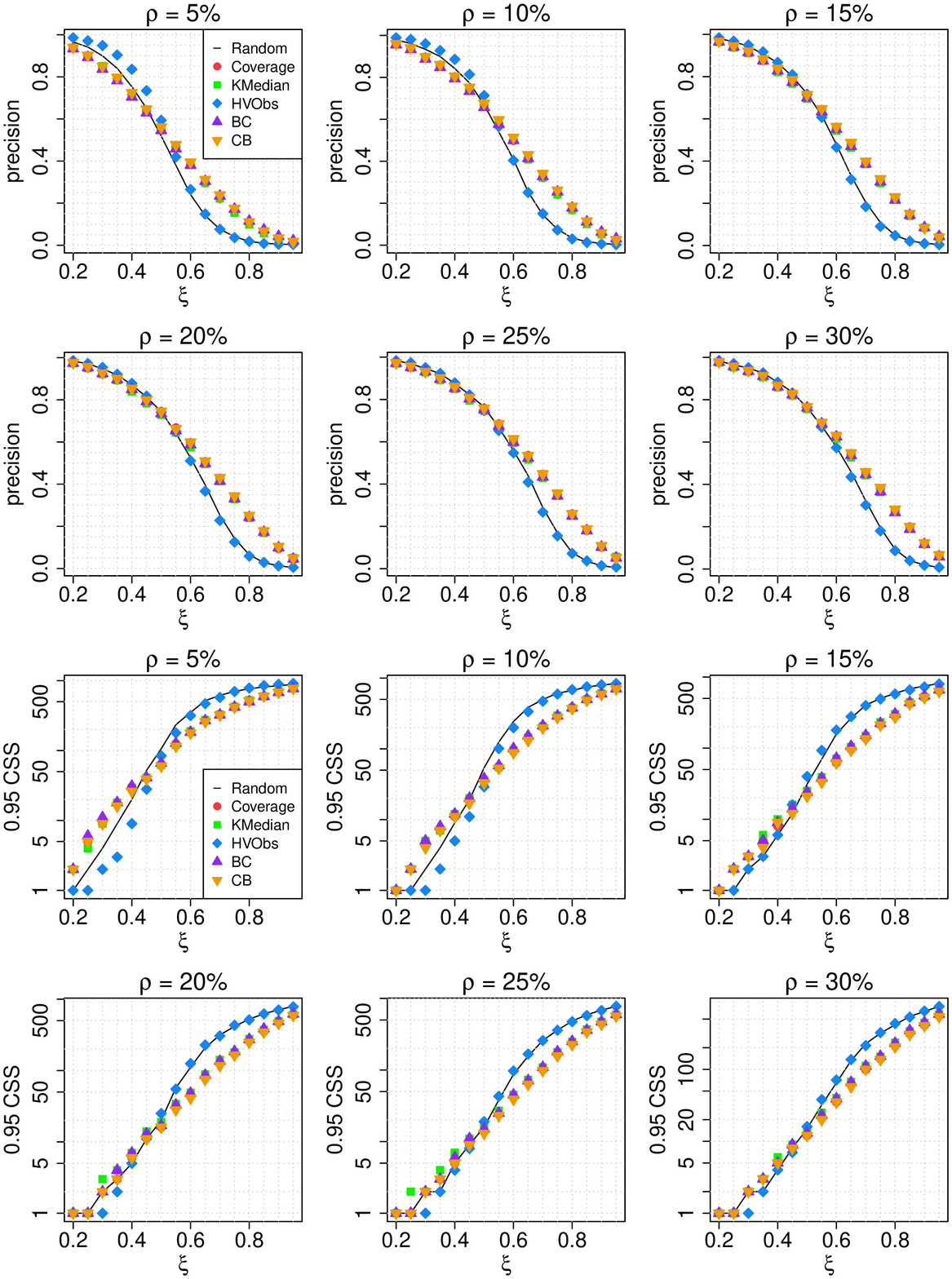}
    \caption{Comparison of source localization quality for five sensor placement strategies in case of \textbf{Barab\'{a}si-Albert model} ($n=1000$, $\langle k \rangle = 7.98$).
             A black solid line denotes the results for randomly placed sensors.
             The confidence intervals at the level $0.95$ are smaller than symbols.}
    \label{fig:ba_results}
\end{figure*}

\begin{figure*}
    \centering
    \includegraphics[width=0.96\textwidth]{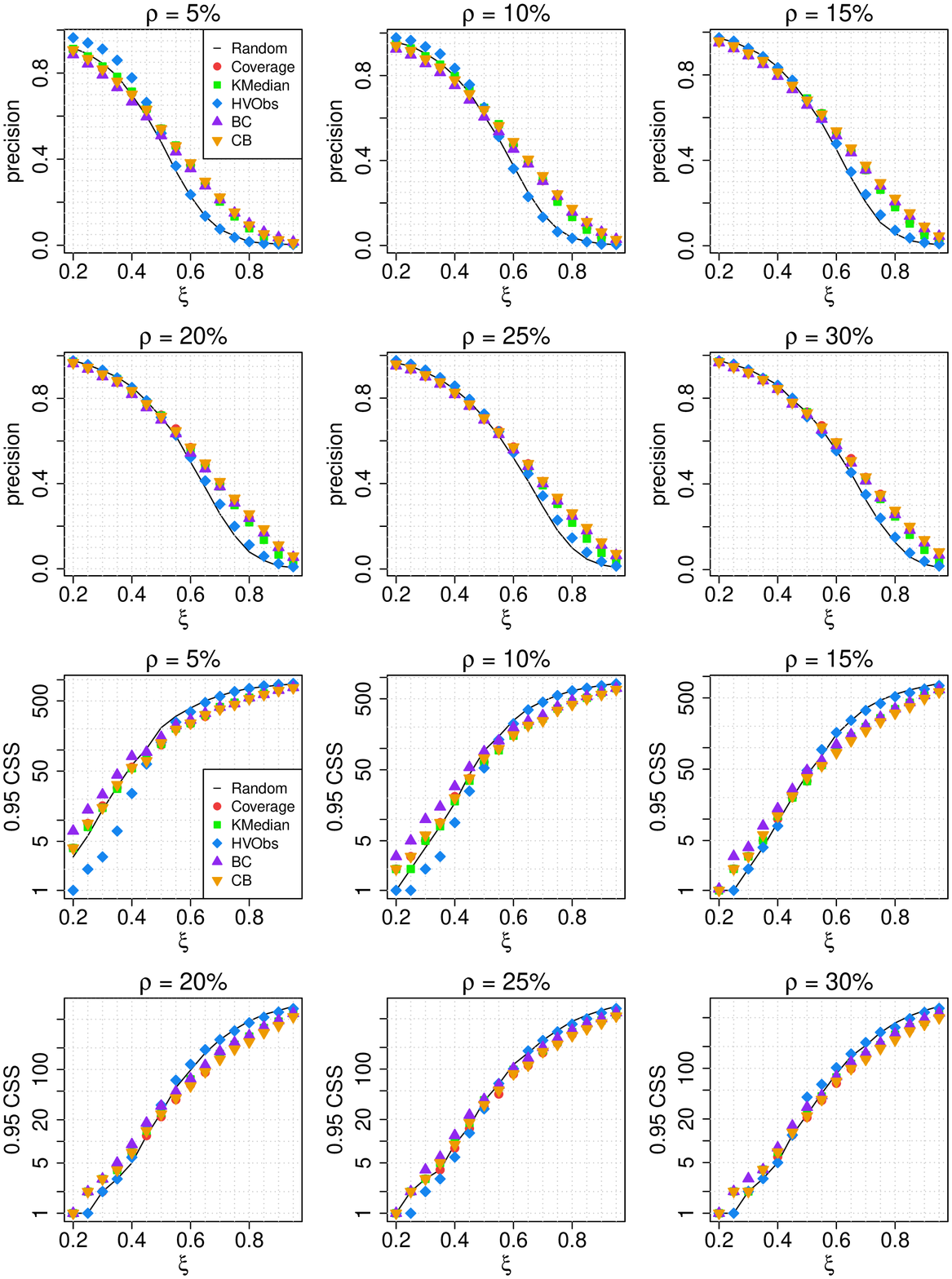}
    \caption{Comparison of source localization quality for five sensor placement strategies in case of \textbf{Configuration Model} ($n=1000$, $\langle k \rangle = 7.02$).
             A black solid line denotes the results for randomly placed sensors.
             The confidence intervals at the level $0.95$ are smaller than symbols.}
    \label{fig:sfn_results}
\end{figure*}

\begin{figure*}
    \centering
    \includegraphics[width=0.96\textwidth]{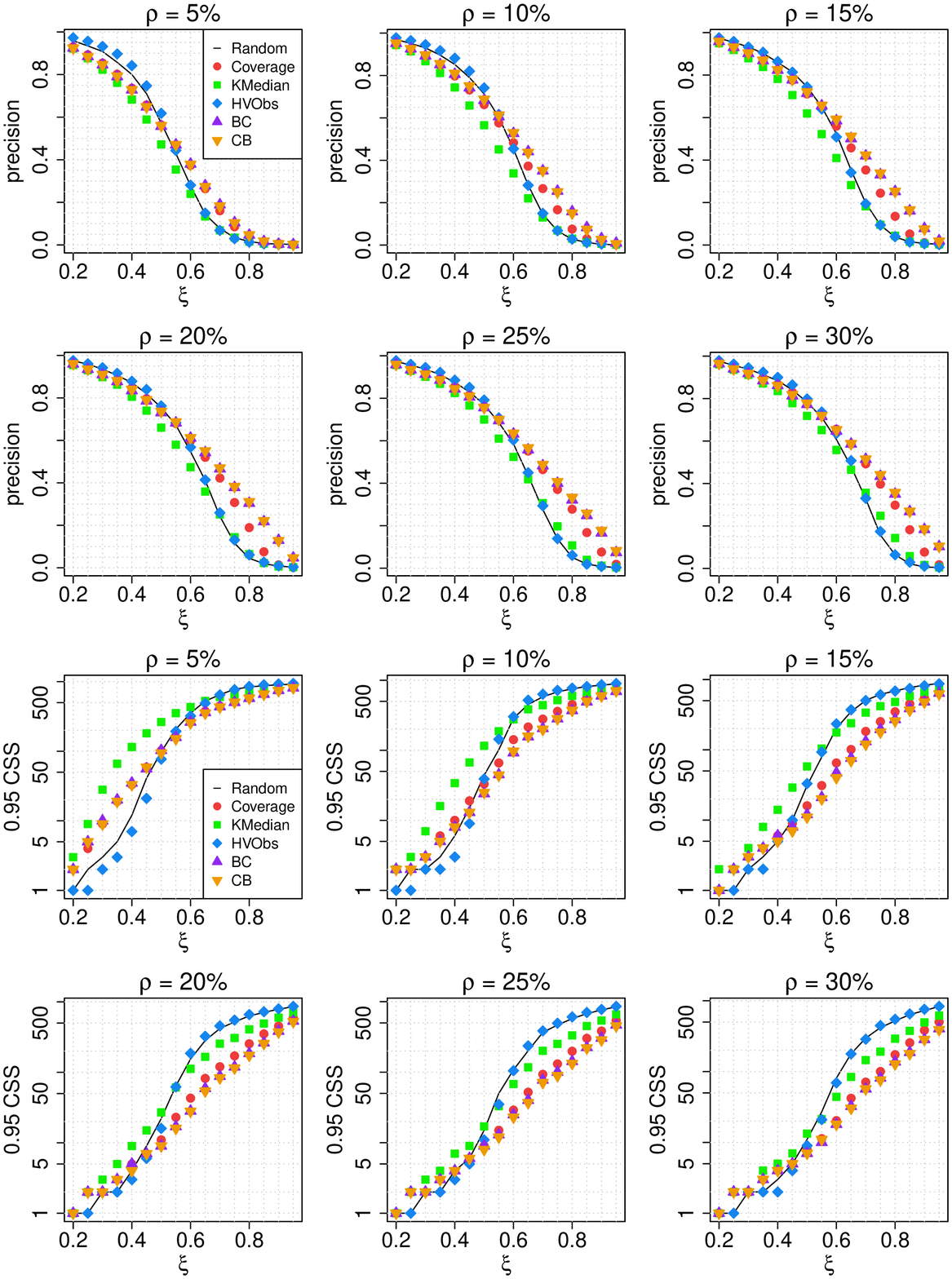}
    \caption{Comparison of source localization quality for five sensor placement strategies in case of \textbf{University of California network} ($n=1020$, $\langle k \rangle = 12$).
             A black solid line denotes the results for randomly placed sensors.
             The confidence intervals at the level $0.95$ are smaller than symbols.}
    \label{fig:uc_results}
\end{figure*}

\begin{figure*}
    \centering
    \includegraphics[width=0.96\textwidth]{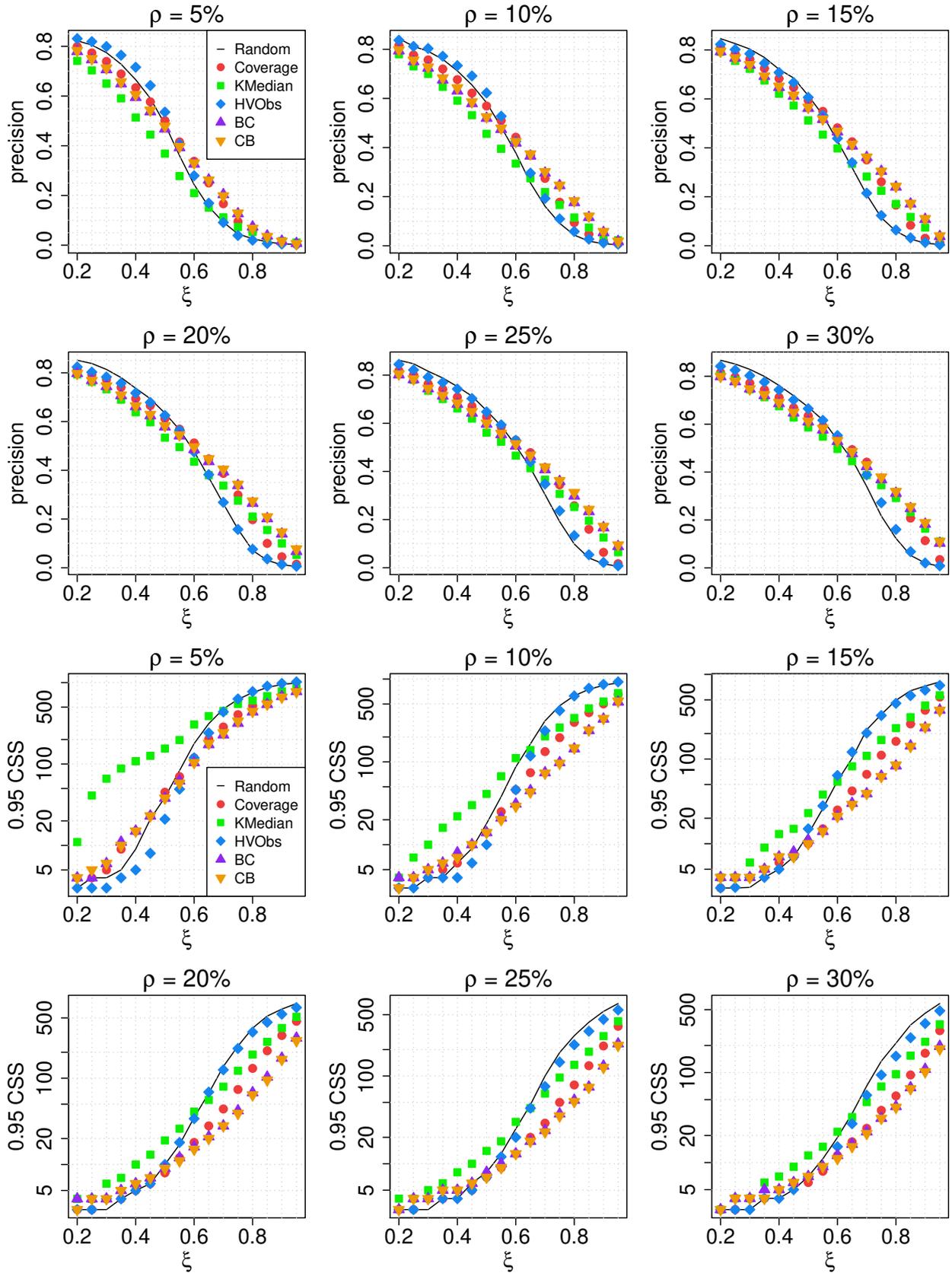}
    \caption{Comparison of source localization quality for five sensor placement strategies in case of \textbf{University of Rovira i Virgili network} ($n=1133$, $\langle k \rangle = 9.6$).
             A black solid line denotes the results for randomly placed sensors.
             The confidence intervals at the level $0.95$ are smaller than symbols.}
    \label{fig:rovira_results}
\end{figure*}

\begin{figure*}
    \centering
    \includegraphics[width=0.96\textwidth]{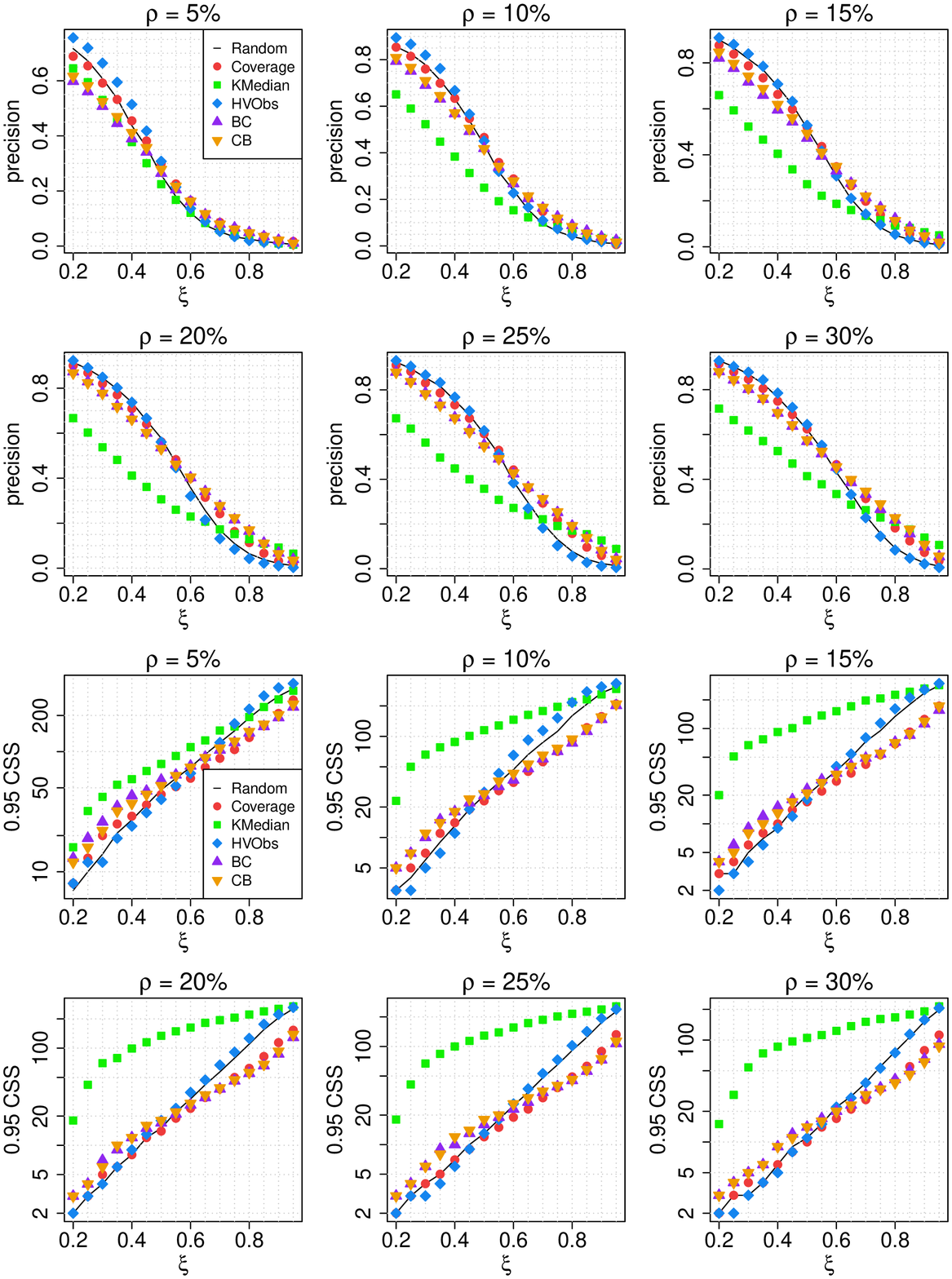}
    \caption{Comparison of source localization quality for five sensor placement strategies in case of \textbf{Infectious network} ($n=410$, $\langle k \rangle = 13.5$).
             A black solid line denotes the results for randomly placed sensors.
             The confidence intervals at the level $0.95$ are smaller than symbols.}
    \label{fig:infect_results}
\end{figure*}

\begin{figure*}
    \centering
    \includegraphics[width=0.96\textwidth]{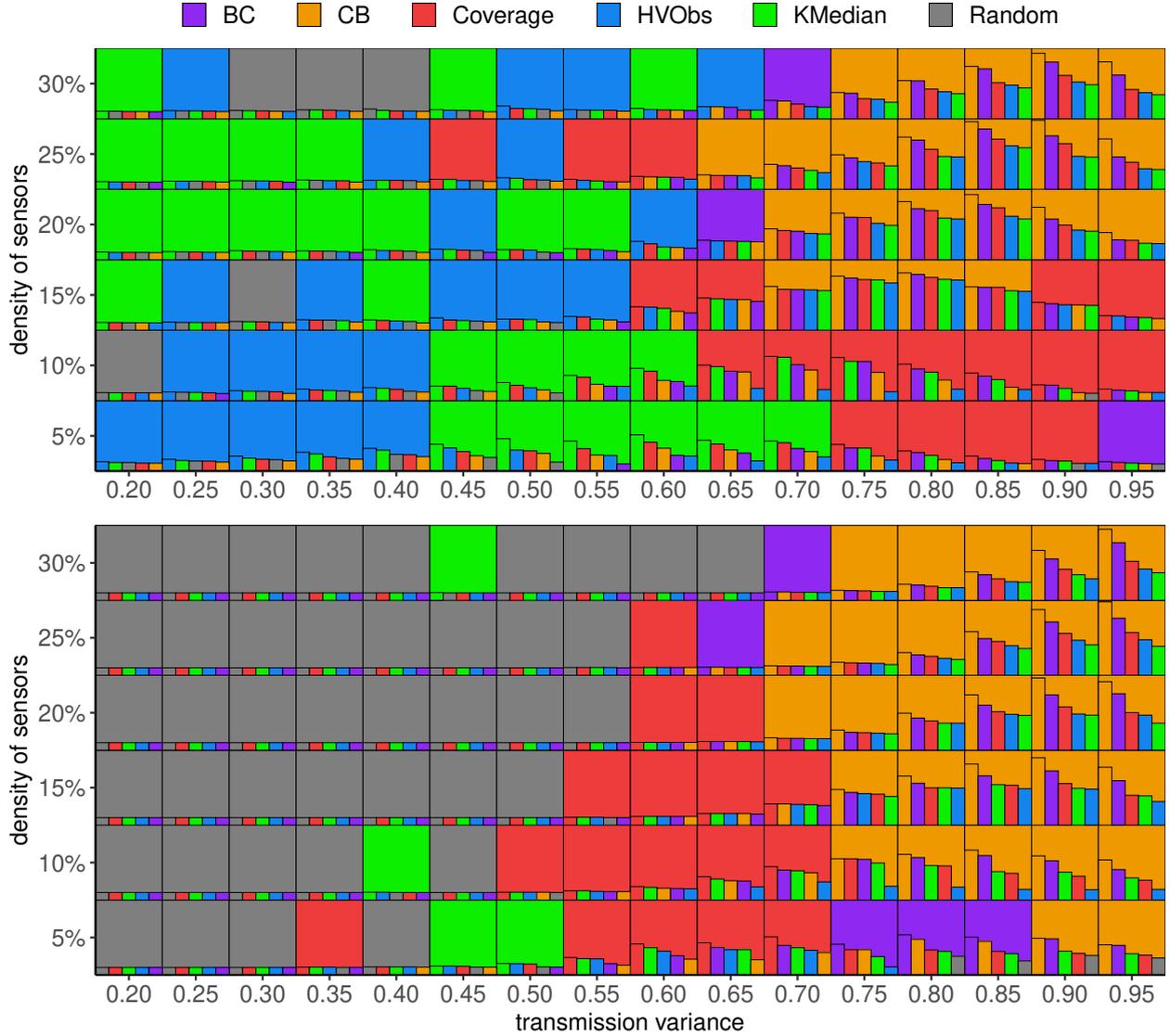}
    \caption{Summary Diagrams for \textbf{Erd\H{o}s-R\'{e}nyi} graph ($n=1000$, $\langle k \rangle = 8$).
             The~color of the background in each tile indicates which algorithm provides the highest average precision (top) or the smallest 0.95-CSS (bottom) for a given pair ($\xi,\rho$).
             The five colored bars inside each tile, ordered from highest to lowest, illustrate the~ranking of~the~methods.
             The height of~each bar shows the difference in average precision (or 0.95-CSS) between a given method and the last method.
             The last (sixth) method, which is the least effective for a given pair ($\xi,\rho$) is not shown in~a~given tile.
             The heights of~bars from all tiles are scaled relative to the height of the highest bar among them, with a minimum height to recognize the color.
             The~bars on the left sides of Summary Diagrams (for low transmission variance $\xi$) are much lower than bars on the right sides, which means that the differences in average precision and 0.95-CSS between methods are much smaller for low values of~$\xi$ than for large $\xi$.
             In case of ER graph, the highest bar in Summary Diagram for average precision (top, $\xi=0.9, \rho=25\%$) represents difference of~18(1) percent points, while the highest bar for 0.95-CSS (bottom, $\xi=0.95, \rho=25\%$) corresponds to 230 nodes.}
    \label{fig:er_diagrams}
\end{figure*}

\begin{figure*}
    \centering
    \includegraphics[width=0.96\textwidth]{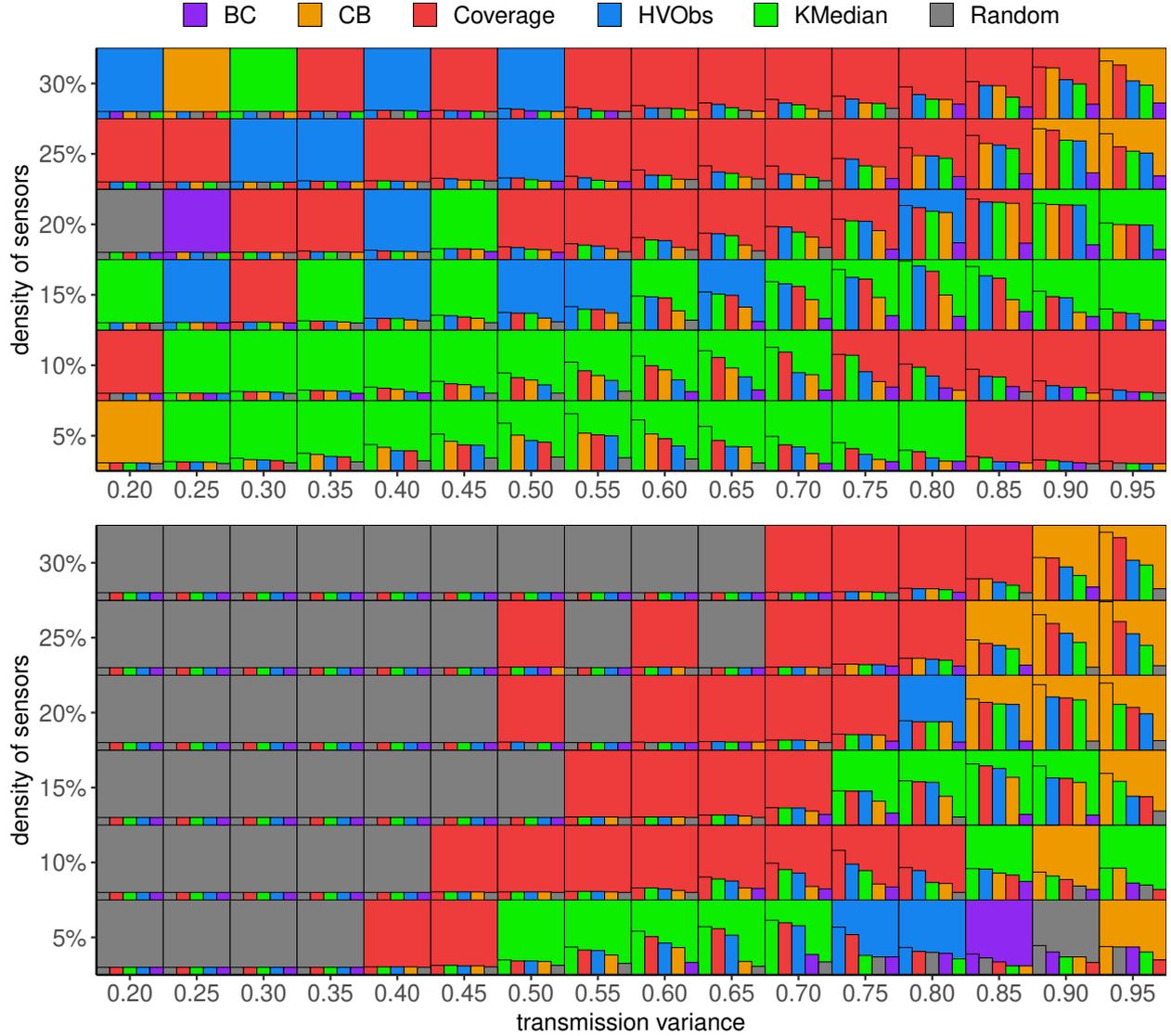}
    \caption{Summary Diagrams for \textbf{Random Regular Graph} ($n=1000$, $\langle k \rangle = 8$).
             The highest bar for average precision (top diagram, $\xi=0.8, \rho=15\%$) represents difference of~15(1) percent points, while for 0.95-CSS the highest bar (bottom diagram, $\xi=0.95, \rho=25\%$) corresponds to 133 nodes.
             See Fig. \ref{fig:er_diagrams} for detailed instruction how to read Summary Diagrams.}
    \label{fig:rrg_diagrams}
\end{figure*}

\begin{figure*}
    \centering
    \includegraphics[width=0.96\textwidth]{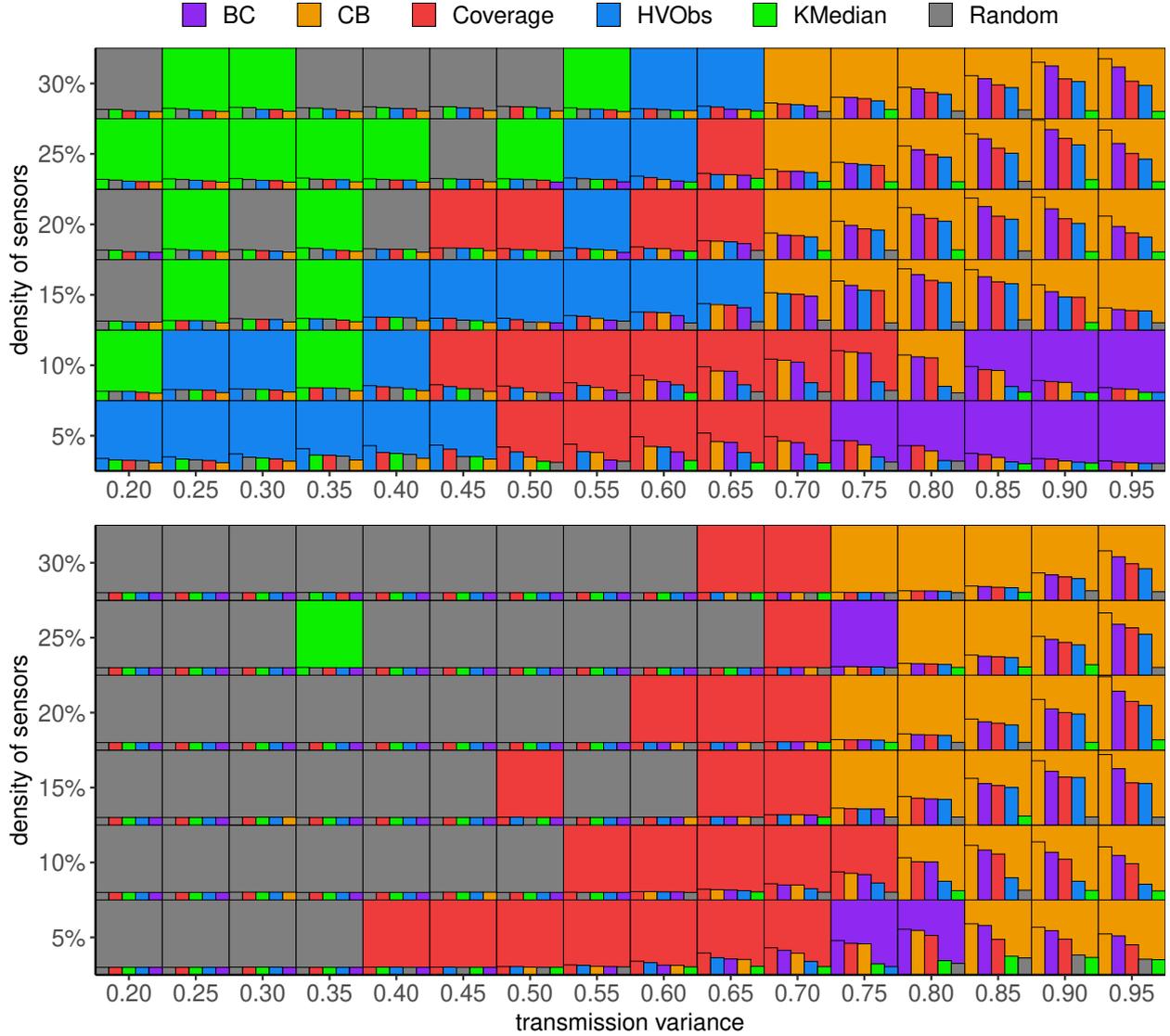}
    \caption{Summary Diagrams for \textbf{Degree Sequence Algorithm} ($n=1000$, $\langle k \rangle = 7.67$).
             The highest bar for average precision (top diagram, $\xi=0.9, \rho=25\%$) represents difference of~19(1) percent points, while for 0.95-CSS the highest bar (bottom diagram, $\xi=0.95, \rho=20\%$) corresponds to 266 nodes.
             See Fig. \ref{fig:er_diagrams} for detailed instruction how to read Summary Diagrams.}
    \label{fig:pois_diagrams}
\end{figure*}

\begin{figure*}
    \centering
    \includegraphics[width=0.96\textwidth]{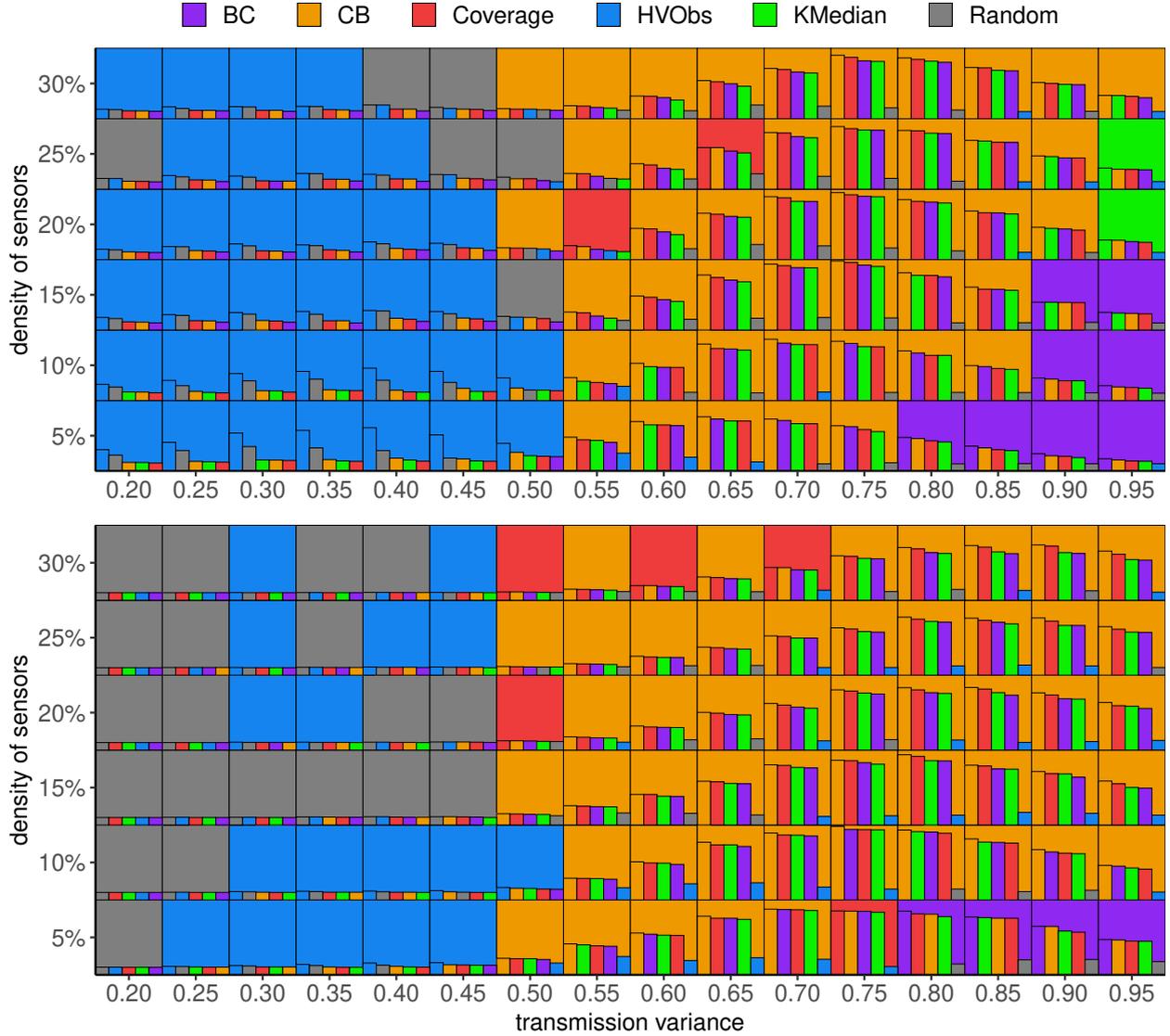}
    \caption{Summary Diagrams for \textbf{Barab\'{a}si-Albert} model ($n=1000$, $\langle k \rangle = 7.98$).
             The highest bar for average precision (top diagram, $\xi=0.75, \rho=15\%$) represents difference of~22(1) percent points, while for 0.95-CSS the highest bar (bottom diagram, $\xi=0.75, \rho=10\%$) corresponds to 334 nodes.
             See Fig. \ref{fig:er_diagrams} for detailed instruction how to read Summary Diagrams.}
    \label{fig:ba_diagrams}
\end{figure*}

\begin{figure*}
    \centering
    \includegraphics[width=0.96\textwidth]{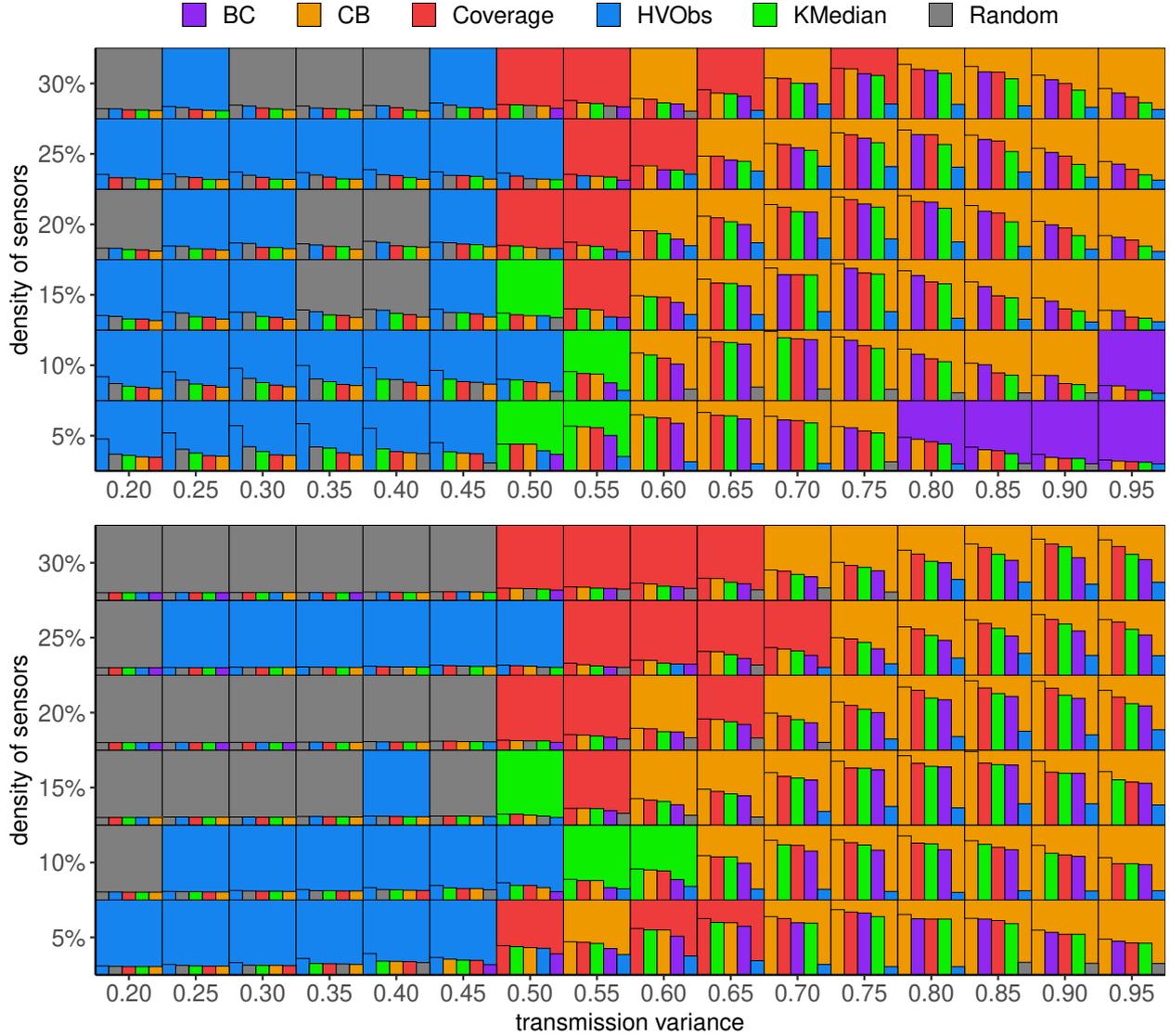}
    \caption{Summary Diagrams for \textbf{Configuration Model} ($n=1000$, $\langle k \rangle = 7.02$).
             The highest bar for average precision (top diagram, $\xi=0.7, \rho=10\%$) represents difference of~19(1) percent points, while for 0.95-CSS the highest bar (bottom diagram, $\xi=0.85, \rho=15\%$) corresponds to 273 nodes.
             See Fig. \ref{fig:er_diagrams} for detailed instruction how to read Summary Diagrams.}
    \label{fig:sfn_diagrams}
\end{figure*}

\begin{figure*}
    \centering
    \includegraphics[width=0.96\textwidth]{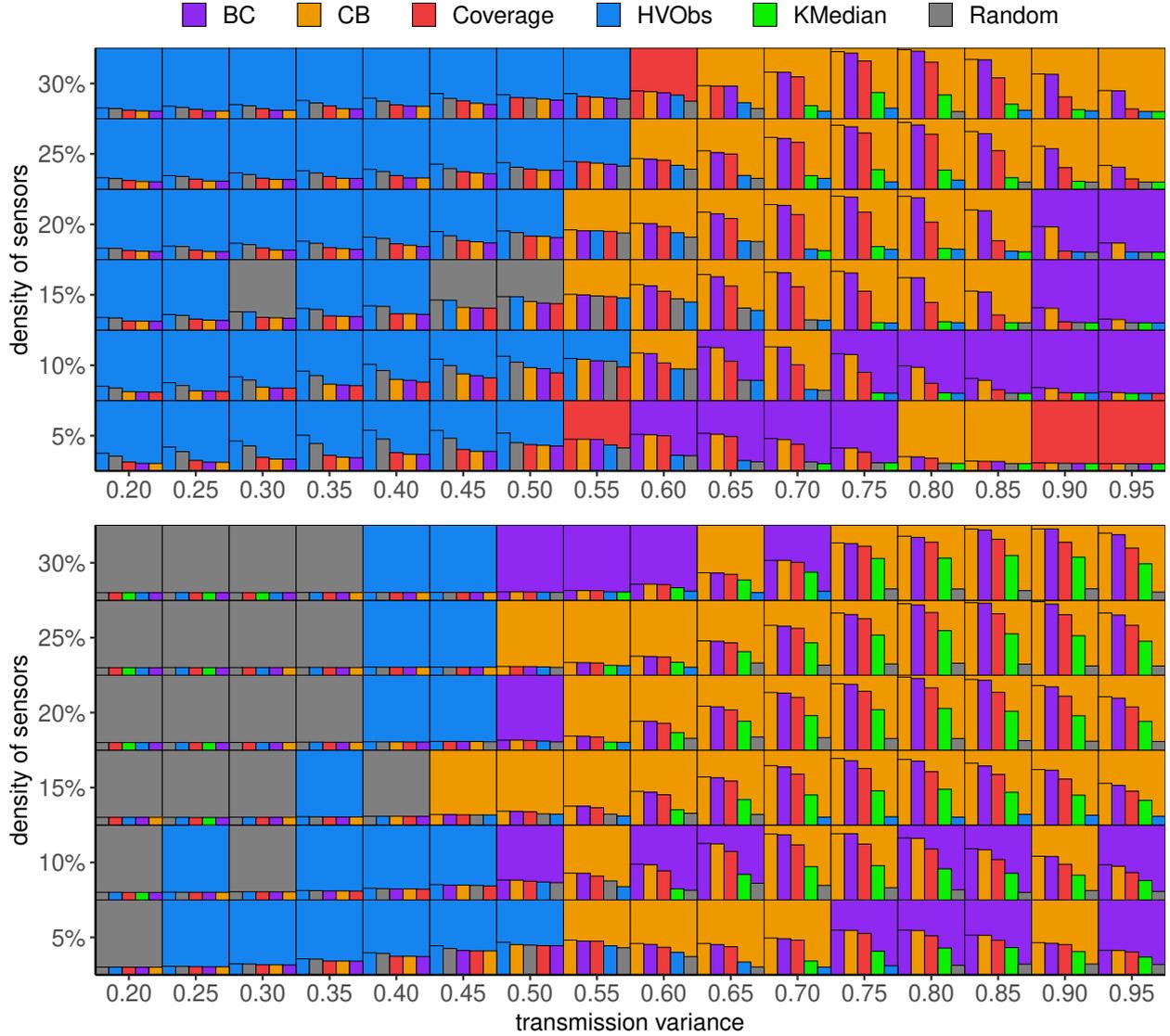}
    \caption{Summary Diagrams for \textbf{University of California network} ($n=1020$, $\langle k \rangle = 12.2$).
             The highest bar for average precision (top diagram, $\xi=0.8, \rho=30\%$) represents difference of~29(1) percent points, while for 0.95-CSS the highest bar (bottom diagram, $\xi=0.9, \rho=25\%$) corresponds to 489 nodes.
             See Fig. \ref{fig:er_diagrams} for detailed instruction how to read Summary Diagrams.}
    \label{fig:uc_diagrams}
\end{figure*}

\begin{figure*}
    \centering
    \includegraphics[width=0.96\textwidth]{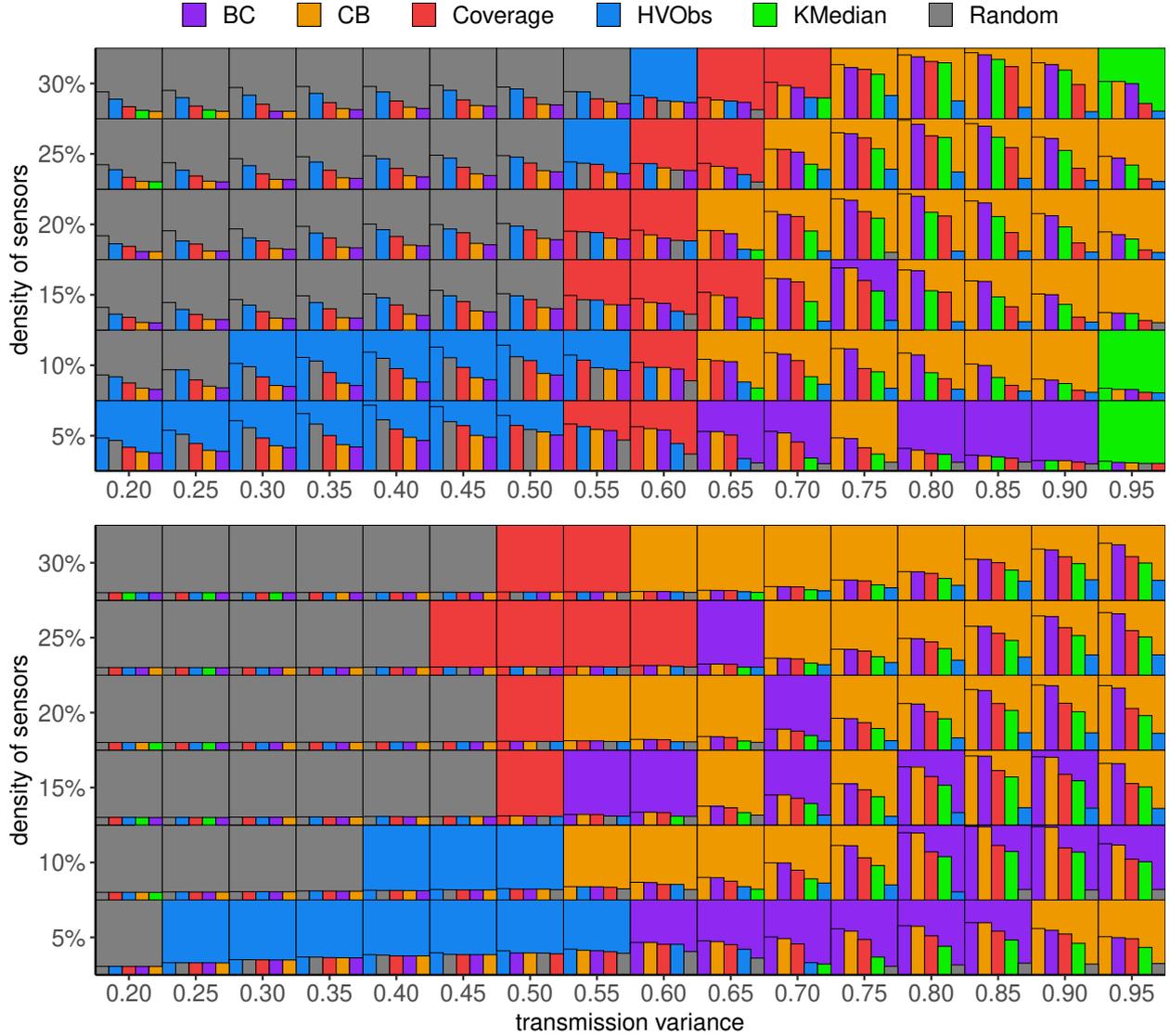}
    \caption{Summary Diagrams for \textbf{University of Rovira i Virgili network} ($n=1133$, $\langle k \rangle = 9.6$).
             The highest bar for average precision (top diagram, $\xi=0.8, \rho=25\%$) represents difference of~21(1) percent points, while for 0.95-CSS the highest bar (bottom diagram, $\xi=0.85, \rho=10\%$) corresponds to 535 nodes.
             See Fig. \ref{fig:er_diagrams} for detailed instruction how to read Summary Diagrams.}
    \label{fig:rovira_diagrams}
\end{figure*}

\begin{figure*}
    \centering
    \includegraphics[width=0.96\textwidth]{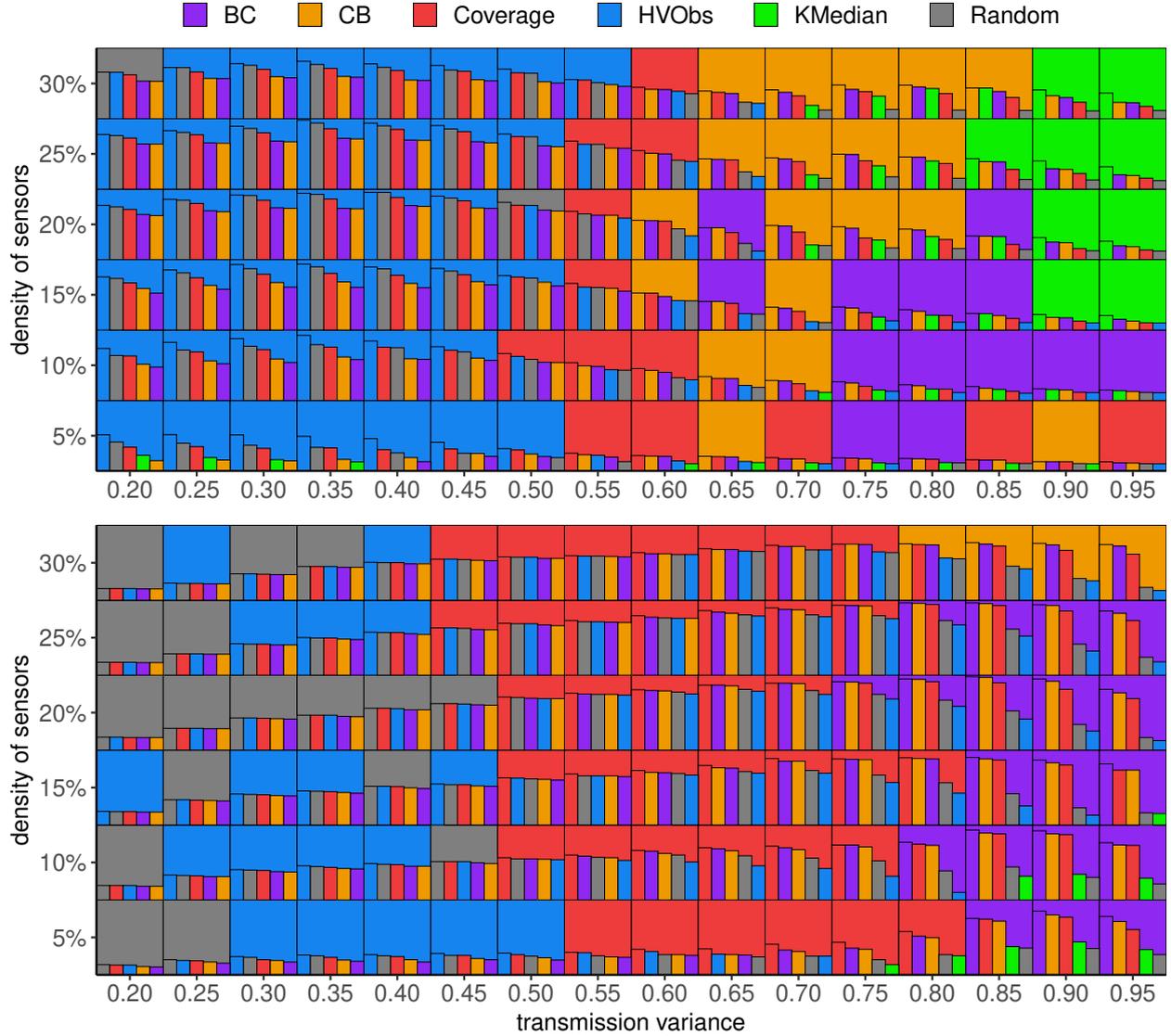}
    \caption{Summary Diagrams for \textbf{Infectious network} ($n=410$, $\langle k \rangle = 13.5$).
             The highest bar for average precision (top diagram, $\xi=0.35, \rho=25\%$) represents difference of~33(1) percent points, while for 0.95-CSS the highest bar (bottom diagram, $\xi=0.85, \rho=20\%$) corresponds to 173 nodes.
             See Fig. \ref{fig:er_diagrams} for detailed instruction how to read Summary Diagrams.}
    \label{fig:infect_diagrams}
\end{figure*}

As shown in Tables \ref{tab:er}-\ref{tab:infect} and in Figures \ref{fig:er_results}-\ref{fig:infect_results}, right choice of sensor placement can significantly increase precision and reduce Credible Set Size. However, the gain of doing so varies from very high values in some cases to moderate in others. For example, for the high values of $\xi$ and $\rho$ (see third column in Tab. \ref{tab:er}-\ref{tab:infect}) the increase of average precision can reach even 18 percent (from 2.1 to 20.2, for the University of California network) in comparison to the baseline random method.
On the other hand, for low values of transmission variance $\xi \in \langle 0.2;0.5 \rangle$, the performance of different sets of sensors is very similar and a noticeable gain can be observed only for the scale-free networks when the density of sensors is rather low.
Figures \ref{fig:er_diagrams}-\ref{fig:infect_diagrams} show which methods outperforms the others for the given transmission variance $\xi$ and density of sensors $\rho$.
These intricate, colorful mosaics can be difficult to interpret but some patterns recur across different networks.
The first thing that catches the eye is the difference between networks with a narrow degree distribution (Erd\H{o}s-R\'{e}nyi, Random Regular Graph and Degree Sequence Algorithm) and scale-free networks. Unlike the former, performance of the later splits the parameter space into two clear parts.
The summary maps for Barab\'{a}si-Albert, Configuration Model, University of California, University of Rovira and Infectious networks are visibly divided into two parts. The left part corresponds to the low transmission variance, and it is dominated by HV-Obs and Random. The right part represents spreading with higher stochasticity. It is more favorable to CB or BC (with the exception of the Infectious network, where K-Median also appears as a leader in a small region).
However, in case of ER, RRG and DSA networks, Coverage and K-Median are also performing quite well along with HV-Obs and CB and the domains of the particular methods are usually more fragmented for these networks than for the real and scale-free networks.

In summary, among all the studied algorithms for sensor placement, two of them perform particularly well.
The first one is High Variance Observers which outperforms the others when transmission variance is low, and the second one is Collective Betweenness which provides the highest quality of source localization when spreading is highly stochastic and unpredictable.
Despite the large number of tests carried out in this review, there is still place for additional research on this topic.
This work has been done for unweighted networks, while the connections in real-world networks are usually not identical and thus future testing of the source localization on a weighted graphs would be of great value.
Similarly, in our study, we perform source location when the signal has already reached all sensors whereas one could imagine situations when we are trying to locate the source as soon as a small fraction of sensors is reached by the spread and under such conditions one could imagine significantly different results than presented here by us.

%%% Plots with average precision and 0.95-CSS

%%% Summary Diagrams and Summary Tables 

\begin{table*}
    \centering
    \caption{\label{tab:er}
             Summary Table for \textbf{Erd\H{o}s-R\'{e}nyi} graph ($n=1000$, $\langle k \rangle = 8$) presents the average values of precision (top numbers in cells, in percentages) and Credible Set Size (bottom numbers in~cells) in~nine regions of~parameter space ($\xi, \rho$).
             The first three numerical columns from the left refer to low, three in the middle to medium, and the last three to high transmission variance $\xi$.
             Similarly, columns $\{1,4,7\}$ correspond to high, $\{2,5,8\}$ to middle, and  $\{3,6,9\}$ to low density of~sensors $\rho$.
             Due to arrangement of the columns, the average precision (0.95-CSS) always decreases (increases) from left to right side of the table.
             The best results in each region are~printed in bold. The uncertainty of average precision is given by the confidence interval at~the~level $0.95$.
             In case of ER graph, K-Median, HV-Obs and Coverage are the best methods for low transmission variance, while Collective Betweenness is the leading method for high values of $\xi$.
             However, Coverage is also doing well for high $\xi$ if the density of sensors $\rho$ is~low.}
    \setlength{\tabcolsep}{4.5pt}
    \begin{tabular}{|r||c|c|c||c|c|c||c|c|c|}
        \hline
        $\xi~\xrightarrow{}$ & \multicolumn{3}{c||}{$\langle 0.2;0.5 \rangle$} & \multicolumn{3}{c||}{$\langle 0.5;0.8 \rangle$} & \multicolumn{3}{c|}{$\langle 0.8;0.95 \rangle$} \\
        \hline
        $\rho~[\%]\xrightarrow{}$ & 20-30 & 10-20 & 5-10 & 20-30 & 10-20 & 5-10 & 20-30 & 10-20 & 5-10 \\ \hline
        \multirow{2}{*}{Random} & 96.9(1) & 95.9(1) & 93.3(1) & 67.3(2) & 57.3(2) & 43.4(3) & 16.6(2) & 8.7(2) & 4.0(1) \\ 
        & 1 & 1 & 2 & 17 & 90 & 282 & 307 & 523 & 705 \\ 
        \hline
        \multirow{2}{*}{Coverage} & \textbf{97.0(1)} & \textbf{96.2(1)} & 93.9(1) & 70.9(2) & \textbf{63.9(2)} & 49.6(3) & 25.7(3) & 16.0(2) & \textbf{7.2(2)} \\ 
        & 1 & 1 & 2 & 8 & \textbf{33} & 191 & 191 & 437 & 682 \\
        \hline
        \multirow{2}{*}{K-Median} & \textbf{97.1(1)} & \textbf{96.3(1)} & \textbf{94.6(1)} & 70.2(2) & 63.4(2) & \textbf{50.0(2)} & 23.5(2) & 14.6(2) & 6.1(2) \\ 
        & 1 & 1 & 2 & 9 & 38 & 214 & 220 & 434 & 682 \\
        \hline
        \multirow{2}{*}{HV-Obs} & \textbf{97.0(1)} & \textbf{96.3(1)} & 94.4(1) & 70.3(2) & 61.6(2) & 45.1(3) & 23.5(3) & 13.9(2) & 4.4(1) \\
        & 1 & 1 & 2 & 9 & 55 & 260 & 219 & 484 & 721 \\ 
        \hline
        \multirow{2}{*}{BC} & 96.5(1) & 95.5(1) & 92.3(1) & 71.1(2) & 62.9(2) & 47.3(3) & 27.9(3) & \textbf{16.1(2)} & 6.7(2) \\
        & 1 & 1 & 2 & 7 & 39 & 198 & 155 & 386 & \textbf{626} \\ 
        \hline
        \multirow{2}{*}{CB} & 96.6(1) & 95.7(1) & 93.3(1) & \textbf{71.5(2)} & 63.0(2) & 47.1(3) & \textbf{30.5(3)} & \textbf{16.1(2)} & 4.9(1) \\ 
        & 1 & 1 & 2 & \textbf{6} & 36 & 221 & \textbf{108} & \textbf{354} & \textbf{625} \\ 
        \hline
    \end{tabular}
\end{table*}

\clearpage

\begin{table*}
    \centering
    \caption{\label{tab:rrg} Summary Table for \textbf{Random Regular Graph} ($n=1000$, $\langle k \rangle = 8$).
             See Table \ref{tab:er} for detailed instruction how to read Summary Table.}
    \setlength{\tabcolsep}{4.5pt}
     \begin{tabular}{|r||c|c|c||c|c|c||c|c|c|}
        \hline
        $\xi~\xrightarrow{}$ & \multicolumn{3}{c||}{$\langle 0.2;0.5 \rangle$} & \multicolumn{3}{c||}{$\langle 0.5;0.8 \rangle$} & \multicolumn{3}{c|}{$\langle 0.8;0.95 \rangle$} \\
        \hline
        $\rho~[\%]\xrightarrow{}$ & 20-30 & 10-20 & 5-10 & 20-30 & 10-20 & 5-10 & 20-30 & 10-20 & 5-10 \\ \hline
        \multirow{2}{*}{Random} & 98.6(1) & 97.7(1) & 95.1(1) & 73.1(2) & 63.1(2) & 48.0(2) & 22.1(2) & 12.0(2) & 5.5(2) \\ 
        & 1 & 1 & 1 & 8 & 48 & 217 & 216 & 439 & \textbf{657} \\ 
        \hline
        \multirow{2}{*}{Coverage} & \textbf{98.9(1)} & 98.4(1) & 96.5(1) & \textbf{76.9(2)} & 69.6(2) & 53.9(3) & \textbf{32.1(3)} & 19.7(2) & \textbf{8.3(2)} \\ 
        & 1 & 1 & 1 & \textbf{4} & \textbf{17} & \textbf{151} & 122 & 395 & 675 \\
        \hline
        \multirow{2}{*}{K-Median} & \textbf{98.8(1)} & \textbf{98.5(1)} & \textbf{97.4(1)} & 75.6(2) & \textbf{70.0(2)} & \textbf{56.0(2)} & 30.1(3) & \textbf{19.8(2)} & 7.6(2) \\ 
        & 1 & 1 & 1 & 5 & 18 & 167 & 144 & \textbf{372} & 669 \\
        \hline
        \multirow{2}{*}{HV-Obs} & \textbf{98.9(1)} & 98.3(1) & 96.3(1) & 76.2(2) & 68.5(2) & 51.7(3) & 30.8(3) & 19.3(2) & 7.1(2) \\ 
        & 1 & 1 & 1 & \textbf{4} & 21 & 169 & 136 & 409 & 671 \\ 
        \hline
        \multirow{2}{*}{BC} & 98.6(1) & 97.6(1) & 94.7(1) & 73.2(2) & 63.5(2) & 48.1(3) & 23.9(2) & 13.5(2) & 6.2(2) \\ 
        & 1 & 1 & 2 & 7 & 43 & 212 & 213 & 434 & \textbf{657} \\ 
        \hline
        \multirow{2}{*}{CB} & 98.7(1) & 98.2(1) & 96.7(1) & 75.0(2) & 66.8(2) & 51.8(3) & 31.7(3) & 16.8(2) & 5.6(2) \\ 
        & 1 & 1 & 1 & 5 & 32 & 212 & \textbf{101} & 378 & 666 \\
        \hline
    \end{tabular}
\end{table*}

\begin{table*}
    \centering
    \caption{\label{tab:pois}
             Summary Table for \textbf{Degree Sequence Algorithm} ($n=1000$, $\langle k \rangle = 7.67$).
             See Table \ref{tab:er} for detailed instruction how to read Summary Table.}
    \setlength{\tabcolsep}{4.5pt}
     \begin{tabular}{|r||c|c|c||c|c|c||c|c|c|}
        \hline
        $\xi~\xrightarrow{}$ & \multicolumn{3}{c||}{$\langle 0.2;0.5 \rangle$} & \multicolumn{3}{c||}{$\langle 0.5;0.8 \rangle$} & \multicolumn{3}{c|}{$\langle 0.8;0.95 \rangle$} \\
        \hline
        $\rho~[\%]\xrightarrow{}$ & 20-30 & 10-20 & 5-10 & 20-30 & 10-20 & 5-10 & 20-30 & 10-20 & 5-10 \\ \hline
        \multirow{2}{*}{Random} & 94.1(1) & 93.1(1) & 91.2(1) & 67.7(2) & 58.6(2) & 45.4(3) & 22.0(2) & 12.1(2) & 5.5(2) \\ 
        & 2 & 2 & 2 & 7 & 29 & 141 & 180 & 402 & 623 \\
        \hline
        \multirow{2}{*}{Coverage} & 93.8(1) & 93.2(1) & 91.9(1) & 70.7(2) & 64.3(2) & \textbf{52.3(3)} & 31.1(3) & 20.1(2) & 9.5(2) \\ 
        & 2 & 2 & 2 & 5 & \textbf{10} & \textbf{75} & 73 & 252 & 545 \\ 
        \hline
        \multirow{2}{*}{K-Median} & \textbf{94.2(1)} & 93.1(1) & 91.2(1) & 67.9(2) & 58.4(2) & 45.2(3) & 22.1(2) & 12.1(2) & 5.6(2) \\ 
        & 2 & 2 & 2 & 7 & 29 & 139 & 180 & 401 & 619 \\ 
        \hline
        \multirow{2}{*}{HV-Obs} & 93.9(1) & \textbf{93.3(1)} & \textbf{92.4(1)} & 70.5(2) & 62.5(2) & 47.9(3) & 29.9(3) & 18.1(2) & 6.3(2) \\ 
        & 2 & 2 & 2 & 5 & 15 & 120 & 86 & 326 & 634 \\ 
        \hline
        \multirow{2}{*}{BC} & 93.1(1) & 92.1(1) & 89.7(2) & 70.4(2) & 63.7(2) & 50.3(2) & 33.6(3) & 21.4(2) & \textbf{9.9(2)} \\ 
        & 2 & 2 & 2 & 5 & 11 & 80 & 59 & 227 & 495 \\ 
        \hline
        \multirow{2}{*}{CB} & 93.3(1) & 92.5(1) & 90.6(2) & \textbf{71.0(2)} & \textbf{64.5(2)} & 50.8(3) & \textbf{35.8(3)} & \textbf{22.8(2)} & 9.2(2) \\ 
        & 2 & 2 & 2 & 5 & \textbf{10} & 85 & \textbf{39} & \textbf{192} & \textbf{489} \\
        \hline
    \end{tabular}
\end{table*}

\begin{table*}
    \centering
    \caption{\label{tab:ba}
             Summary Table for \textbf{Barab\'{a}si-Albert} model ($n=1000$, $\langle k \rangle = 7.98$).
             See Table \ref{tab:er} for detailed instruction how to read Summary Table.}
    \setlength{\tabcolsep}{4.5pt}
     \begin{tabular}{|r||c|c|c||c|c|c||c|c|c|}
        \hline
        $\xi~\xrightarrow{}$ & \multicolumn{3}{c||}{$\langle 0.2;0.5 \rangle$} & \multicolumn{3}{c||}{$\langle 0.5;0.8 \rangle$} & \multicolumn{3}{c|}{$\langle 0.8;0.95 \rangle$} \\
        \hline
        $\rho~[\%]\xrightarrow{}$ & 20-30 & 10-20 & 5-10 & 20-30 & 10-20 & 5-10 & 20-30 & 10-20 & 5-10 \\ \hline
        \multirow{2}{*}{Random} & 89.7(1) & 88.0(1) & 83.1(2) & 42.0(2) & 35.2(2) & 25.3(2) & 3.3(1) & 2.2(1) & 1.2(1) \\ 
        & 4 & 6 & 14 & 264 & 398 & 540 & 673 & 745 & 815 \\ 
        \hline
        \multirow{2}{*}{Coverage} & 88.6(1) & 86.0(1) & 80.2(2) & 51.9(2) & 46.4(2) & 36.2(3) & 15.1(2) & 11.6(2) & 7.0(2) \\ 
        & 4 & 7 & 16 & 97 & 143 & 248 & 434 & 523 & 639 \\ 
        \hline
        \multirow{2}{*}{K-Median} & 87.7(1) & 85.3(1) & 80.4(2) & 50.6(2) & 45.5(2) & 36.2(3) & 15.0(2) & 11.6(2) & 6.7(2) \\ 
        & 5 & 8 & 16 & 105 & 148 & 248 & 453 & 528 & 638 \\ 
        \hline
        \multirow{2}{*}{HV-Obs} & \textbf{89.8(1)} & \textbf{89.3(1)} & \textbf{87.5(2)} & 40.5(2) & 34.7(2) & 26.7(2) & 3.2(1) & 2.0(1) & 1.1(1) \\ 
        & 4 & \textbf{4} & \textbf{6} & 276 & 394 & 531 & 667 & 749 & 834 \\ 
        \hline
        \multirow{2}{*}{BC} & 88.1(1) & 85.3(2) & 79.5(2) & 51.0(2) & 45.8(2) & 36.5(3) & 14.7(2) & 11.9(2) & \textbf{7.9(2)} \\ 
        & 5 & 8 & 17 & 107 & 152 & 248 & 459 & 529 & 624 \\ 
        \hline
        \multirow{2}{*}{CB} & 88.6(1) & 86.2(1) & 80.8(2) & \textbf{52.2(2)} & \textbf{47.2(2)} & \textbf{37.7(3)} & \textbf{15.6(2)} & \textbf{12.3(2)} & \textbf{7.7(2)} \\ 
        & 4 & 7 & 15 & \textbf{93} & \textbf{136} & \textbf{238} & \textbf{422} & \textbf{508} & \textbf{623} \\ 
        \hline
    \end{tabular}
\end{table*}

\begin{table*}
    \centering
    \caption{\label{tab:sfn}
             Summary Table for \textbf{Configuration Model} ($n=1000$, $\langle k \rangle = 7.02$).
             See Table \ref{tab:er} for detailed instruction how to read Summary Table.}
    \setlength{\tabcolsep}{4.5pt}
     \begin{tabular}{|r||c|c|c||c|c|c||c|c|c|}
        \hline
        $\xi~\xrightarrow{}$ & \multicolumn{3}{c||}{$\langle 0.2;0.5 \rangle$} & \multicolumn{3}{c||}{$\langle 0.5;0.8 \rangle$} & \multicolumn{3}{c|}{$\langle 0.8;0.95 \rangle$} \\
        \hline
        $\rho~[\%]\xrightarrow{}$ & 20-30 & 10-20 & 5-10 & 20-30 & 10-20 & 5-10 & 20-30 & 10-20 & 5-10 \\ \hline
        \multirow{2}{*}{Random} & 87.1(1) & 85.5(1) & 78.6(2) & 40.9(2) & 33.9(2) & 23.7(2) & 4.4(1) & 2.5(1) & 1.3(1) \\ 
        & 6 & 8 & 32 & 242 & 355 & 520 & 638 & 714 & 800 \\ 
        \hline
        \multirow{2}{*}{Coverage} & 86.7(1) & 84.7(1) & 78.5(2) & 49.3(2) & 44.0(2) & 34.5(3) & 14.3(2) & 9.8(2) & 5.6(2) \\ 
        & 6 & 8 & 26 & 135 & 181 & 310 & 434 & 523 & 653 \\ 
        \hline
        \multirow{2}{*}{K-Median} & 86.3(1) & 85.0(1) & 79.2(2) & 47.7(2) & 43.6(2) & 34.4(3) & 12.0(2) & 8.8(1) & 5.1(1) \\ 
        & 7 & 8 & 25 & 151 & 189 & 315 & 459 & 528 & 654 \\ 
        \hline
        \multirow{2}{*}{HV-Obs} & \textbf{87.6(1)} & \textbf{86.5(1)} & \textbf{83.3(2)} & 43.1(2) & 35.4(2) & 24.1(2) & 6.3(1) & 3.3(1) & 1.2(1) \\ 
        & \textbf{5} & \textbf{7} & \textbf{14} & 230 & 349 & 511 & 587 & 694 & 803 \\ 
        \hline
        \multirow{2}{*}{BC} & 85.2(1) & 82.4(1) & 75.4(2) & 47.9(2) & 43.0(2) & 33.6(2) & 15.0(2) & 11.6(2) & \textbf{7.1(2)} \\ 
        & 8 & 13 & 38 & 162 & 201 & 333 & 481 & 535 & 651 \\  
        \hline
        \multirow{2}{*}{CB} & 86.0(1) & 84.1(1) & 78.1(2) & \textbf{49.4(2)} & \textbf{45.1(2)} & \textbf{35.7(2)} & \textbf{16.4(2)} & \textbf{12.6(2)} & \textbf{7.0(2)} \\ 
        & 6 & 9 & 27 & \textbf{132} & \textbf{169} & \textbf{304} & \textbf{413} & \textbf{483} & \textbf{630} \\ 
        \hline
    \end{tabular}
\end{table*}

\begin{table*}
    \centering
    \caption{\label{tab:uc}
             Summary Table for \textbf{University of California network} ($n=1020$, $\langle k \rangle = 12$).
             See Table \ref{tab:er} for detailed instruction how to read Summary Table.}
    \setlength{\tabcolsep}{4.5pt}
     \begin{tabular}{|r||c|c|c||c|c|c||c|c|c|}
        \hline
        $\xi~\xrightarrow{}$ & \multicolumn{3}{c||}{$\langle 0.2;0.5 \rangle$} & \multicolumn{3}{c||}{$\langle 0.5;0.8 \rangle$} & \multicolumn{3}{c|}{$\langle 0.8;0.95 \rangle$} \\
        \hline
        $\rho~[\%]\xrightarrow{}$ & 20-30 & 10-20 & 5-10 & 20-30 & 10-20 & 5-10 & 20-30 & 10-20 & 5-10 \\ \hline
        \multirow{2}{*}{Random} & 89.4(1) & 88.1(1) & 84.6(2) & 41.7(2) & 36.5(2) & 27.5(2) & 2.1(1) & 1.5(1) & 0.9(1) \\ 
        & \textbf{3} & 4 & 8 & 359 & 493 & 612 & 757 & 818 & 872 \\
        \hline
        \multirow{2}{*}{Coverage} & 87.9(2) & 85.1(2) & 80.3(2) & 52.5(2) & 43.4(2) & 32.6(2) & 11.0(2) & 4.9(1) & 2.2(1) \\ 
        & \textbf{3} & 6 & 17 & 83 & 178 & 349 & 419 & 567 & 713 \\ 
        \hline
        \multirow{2}{*}{K-Median} & 84.8(2) & 81.1(2) & 75.9(2) & 39.9(2) & 30.7(2) & 22.3(2) & 3.7(1) & 1.8(1) & 1.0(1) \\ 
        & 6 & 14 & 58 & 188 & 337 & 511 & 557 & 676 & 775 \\ 
        \hline
        \multirow{2}{*}{HV-Obs} & \textbf{90.4(1)} & \textbf{89.1(1)} & \textbf{87.3(2)} & 43.2(2) & 36.9(2) & 28.2(2) & 2.5(1) & 1.7(1) & 0.8(1) \\ 
        & \textbf{3} & \textbf{3} & \textbf{5} & 389 & 526 & 626 & 769 & 818 & 887 \\ 
        \hline
        \multirow{2}{*}{BC} & 87.0(2) & 85.0(1) & 80.2(2) & 54.7(2) & 48.4(2) & \textbf{36.0(2)} & 19.6(3) & \textbf{12.2(2)} & \textbf{4.4(1)} \\ 
        & 4 & 5 & 15 & 59 & 125 & 306 & 329 & 499 & 678 \\  
        \hline
        \multirow{2}{*}{CB} & 87.2(2) & 85.3(1) & 80.5(2) & \textbf{55.3(2)} & \textbf{48.7(2)} & \textbf{36.0(2)} & \textbf{20.2(3)} & \textbf{12.1(2)} & 4.1(1) \\ 
        & 4 & 5 & 15 & \textbf{55} & \textbf{122} & \textbf{305} & \textbf{318} & \textbf{498} & \textbf{673} \\ 
        \hline
    \end{tabular}
\end{table*}

\begin{table*}
    \centering
    \setlength{\tabcolsep}{4.5pt}
     \begin{tabular}{|r||c|c|c||c|c|c||c|c|c|}
        \hline
        $\xi~\xrightarrow{}$ & \multicolumn{3}{c||}{$\langle 0.2;0.5 \rangle$} & \multicolumn{3}{c||}{$\langle 0.5;0.8 \rangle$} & \multicolumn{3}{c|}{$\langle 0.8;0.95 \rangle$} \\
        \hline
        $\rho~[\%]\xrightarrow{}$ & 20-30 & 10-20 & 5-10 & 20-30 & 10-20 & 5-10 & 20-30 & 10-20 & 5-10 \\ \hline
        \multirow{2}{*}{Random} & \textbf{77.5(1)} & \textbf{75.2(2)} & 71.6(2) & 39.1(2) & 32.6(2) & 24.3(2) & 4.0(1) & 2.5(1) & 1.5(1) \\ 
        & \textbf{4} & 5 & 7 & 115 & 232 & 399 & 566 & 738 & 880 \\
        \hline
        \multirow{2}{*}{Coverage} & 73.2(2) & 71.7(2) & 68.9(2) & 45.7(2) & 39.6(2) & 30.4(2) & 12.4(2) & 6.9(1) & 3.3(1) \\ 
        & \textbf{4} & 5 & 8 & 32 & 76 & 208 & 255 & 419 & 630 \\
        \hline
        \multirow{2}{*}{K-Median} & 69.2(2) & 66.0(2) & 60.3(2) & 41.1(2) & 33.6(2) & 22.9(2) & 16.1(2) & 9.9(2) & 4.6(1) \\ 
        & 7 & 12 & 57 & 58 & 119 & 313 & 299 & 449 & 670 \\
        \hline
        \multirow{2}{*}{HV-Obs} & 75.5(1) & 74.3(2) & \textbf{74.1(2)} & 40.8(2) & 33.9(2) & 26.9(2) & 5.0(1) & 2.9(1) & 1.7(1) \\ 
        & \textbf{4} & \textbf{4} & \textbf{4} & 99 & 208 & 371 & 480 & 707 & 899 \\
        \hline
        \multirow{2}{*}{BC} & 70.6(2) & 68.5(2) & 65.4(2) & 45.2(2) & 40.1(2) & 31.1(2) & 19.1(2) & 13.5(2) & 6.2(1) \\ 
        & 5 & 6 & 9 & 23 & 40 & 138 & \textbf{137} & \textbf{251} & \textbf{545} \\ 
        \hline
        \multirow{2}{*}{CB} & 70.8(2) & 68.8(2) & 66.1(2) & \textbf{45.9(2)} & \textbf{40.7(2)} & \textbf{31.4(2)} & \textbf{19.9(2)} & \textbf{14.0(2)} & \textbf{6.3(2)} \\ 
        & 5 & 6 & 9 & \textbf{22} & \textbf{39} & 143 & \textbf{128} & \textbf{252} & 549 \\  
        \hline
    \end{tabular}
    \caption{\label{tab:rovira}
             Summary Table for \textbf{University of Rovira i Virgili network} ($n=1133$, $\langle k \rangle = 9.6$).
             See detailed instruction how to read Summary Table under Table \ref{tab:er}.}
\end{table*}

\begin{table*}
    \centering
    \caption{\label{tab:infect}
             Summary Table for \textbf{Infectious network} ($n=410$, $\langle k \rangle = 13.5$).
             See Table \ref{tab:er} for detailed instruction how to read Summary Table.}
    \setlength{\tabcolsep}{4.5pt}
    \begin{tabular}{|r||c|c|c||c|c|c||c|c|c|}
        \hline
        $\xi~\xrightarrow{}$ & \multicolumn{3}{c||}{$\langle 0.2;0.5 \rangle$} & \multicolumn{3}{c||}{$\langle 0.5;0.8 \rangle$} & \multicolumn{3}{c|}{$\langle 0.8;0.95 \rangle$} \\
        \hline
        $\rho~[\%]\xrightarrow{}$ & 20-30 & 10-20 & 5-10 & 20-30 & 10-20 & 5-10 & 20-30 & 10-20 & 5-10 \\ \hline
        \multirow{2}{*}{Random} & 78.8(2) & 73.2(2) & 59.7(3) & 31.6(2) & 24.1(2) & 14.9(2) & 4.0(1) & 2.8(1) & 1.9(1) \\ 
        & \textbf{6} & \textbf{9} & 19 & 51 & 76 & 107 & 181 & 234 & 272 \\ 
        \hline
        \multirow{2}{*}{Coverage} & 77.1(2) & 71.7(2) & 59.8(3) & 36.5(2) & 28.1(2) & \textbf{18.8(2)} & 8.2(2) & 4.5(1) & 3.1(1) \\ 
        & \textbf{6} & 10 & 20 & \textbf{30} & \textbf{44} & \textbf{70} & 100 & 136 & 184 \\
        \hline
        \multirow{2}{*}{K-Median} & 52.0(2) & 46.6(2) & 44.9(3) & 25.4(2) & 17.2(2) & 12.0(2) & 13.1(2) & 7.1(1) & 2.8(1) \\ 
        & 84 & 84 & 68 & 167 & 178 & 155 & 228 & 256 & 265 \\ 
        \hline
        \multirow{2}{*}{HV-Obs} & \textbf{79.8(2)} & \textbf{74.9(2)} & \textbf{64.3(2)} & 30.3(2) & 23.6(2) & 16.1(2) & 2.8(1) & 2.5(1) & 1.9(1) \\ 
        & \textbf{6} & \textbf{9} & \textbf{17} & 56 & 94 & 120 & 202 & 271 & 311 \\
        \hline
        \multirow{2}{*}{BC} & 72.7(2) & 66.4(2) & 53.2(2) & 36.8(2) & 28.5(2) & 18.2(2) & 11.4(2) & \textbf{7.3(1)} & \textbf{4.1(1)} \\ 
        & 9 & 13 & 27 & 31 & 46 & 81 & \textbf{77} & \textbf{122} & \textbf{177} \\ 
        \hline
        \multirow{2}{*}{CB} & 72.6(2) & 67.4(2) & 54.7(3) & \textbf{37.2(2)} & \textbf{28.8(2)} & 18.4(2) & \textbf{11.7(2)} & 6.8(1) & 3.7(1) \\ 
        & 9 & 13 & 24 & 32 & 48 & 82 & 79 & 130 & 185 \\  
        \hline
    \end{tabular}
\end{table*}

\clearpage

\printcredits

\section*{Declaration of competing interest}
The authors declare that they have no known competing financial interests or personal relationships that could have appeared to influence the work reported in this paper.

\section*{Acknowledgments}
The work was partially supported as RENOIR Project by the European Union Horizon 2020 Research and Innovation Programme under the Marie Sk\l odowska-Curie grant agreement No 691152 and by Ministry of Science and Higher Education (Poland), Grant Nos. 34/H2020/2016, 329025/\allowbreak PnH/\allowbreak 2016 and the National Science Centre, Poland Grant No. 2015/\allowbreak 19/\allowbreak B/\allowbreak ST6/02612.
R.P. was partially supported by the National Science Centre, Poland, agreement No UMO-2019/32/T/\allowbreak ST6/\allowbreak 00173, and by PLGrid Infrastructure. 
J.A.H. was partially supported by the Russian Scientific Foundation, Agreement \#17-71-30029 with co-financing of Bank Saint Petersburg.
B.K.S. was partially supported by the U.S. Army Research Office Grant No. W911NF-16-1-0524, and by the National Science Centre, Poland Grant No. 2016/\allowbreak 21/\allowbreak B/\allowbreak ST6/01463.

\bibliographystyle{elsarticle-num}
\bibliography{main_bibliography}

\bio{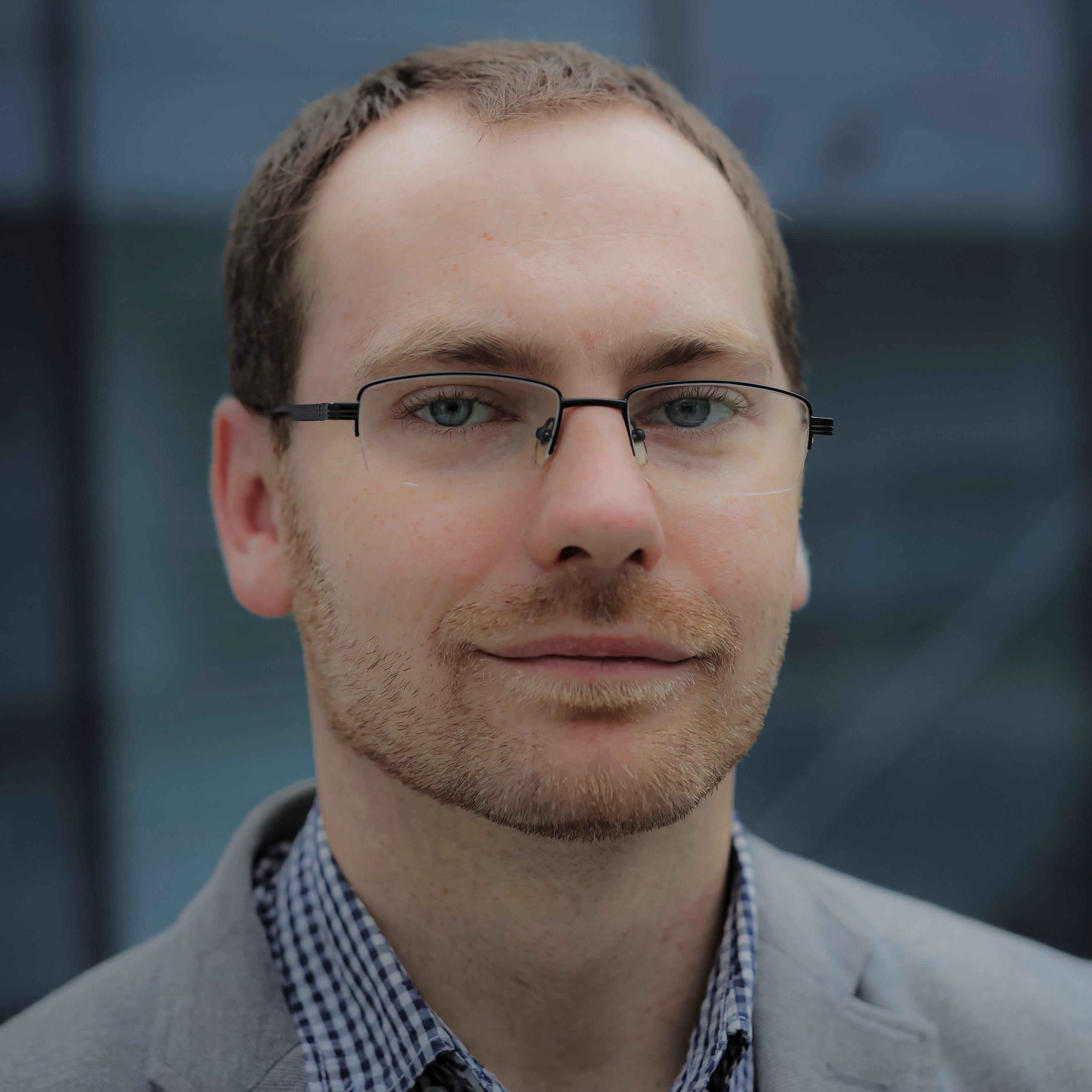}
Robert Paluch received his B.Sc.Eng degree in~Computer Physics in 2012 and M.Sc.Eng in~Complex Systems Modeling in 2014 from Warsaw University of Technology, Poland.
In 2012-2013 he worked for The European Organization for Nuclear Research (CERN) as a participant of the Technical Student Programme.
Currently he is a~PhD student at the Warsaw University of Technology in the Group of Physics in Economy and Social Sciences.
In years 2016-2019 he was visiting researcher at Rensselaer Polytechnic Institute.
\endbio

\bio{}
Łukasz G. Gajewski is a PhD student at the Warsaw University of Technology in the Group of Physics in Economy and Social Sciences. He has received his B.Sc.Eng degree in Computer Physics and M.Sc. Eng~in~Complex Systems Modeling. In years 2018 - 2019 he seconded at Stanford University and Rensselaer Polytechnic Institute.
\endbio

\bio{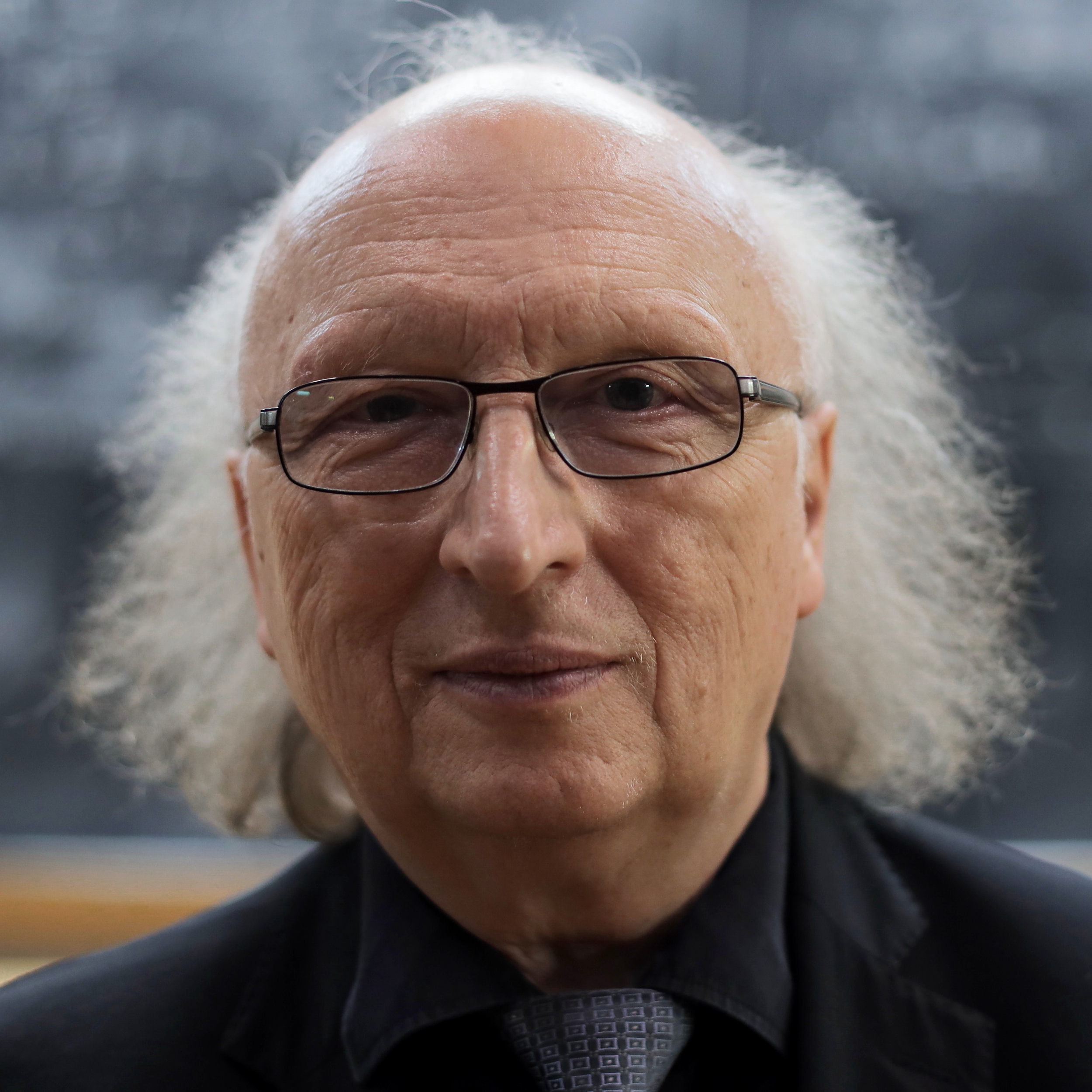}
Janusz A. Holyst is a Full Professor at Faculty of Physics, Warsaw University of Technology where he leads Group of Physics in Economy and Social Sciences.
His current research includes simulations of evolving networks, models of collective opinion and emotion formation, and phase transitions. His list of publications includes around 150 papers in peer reviewed journals that have been cited almost 3000 times. 
He is currently one of Main Editors of Physica A.
\endbio

\bio{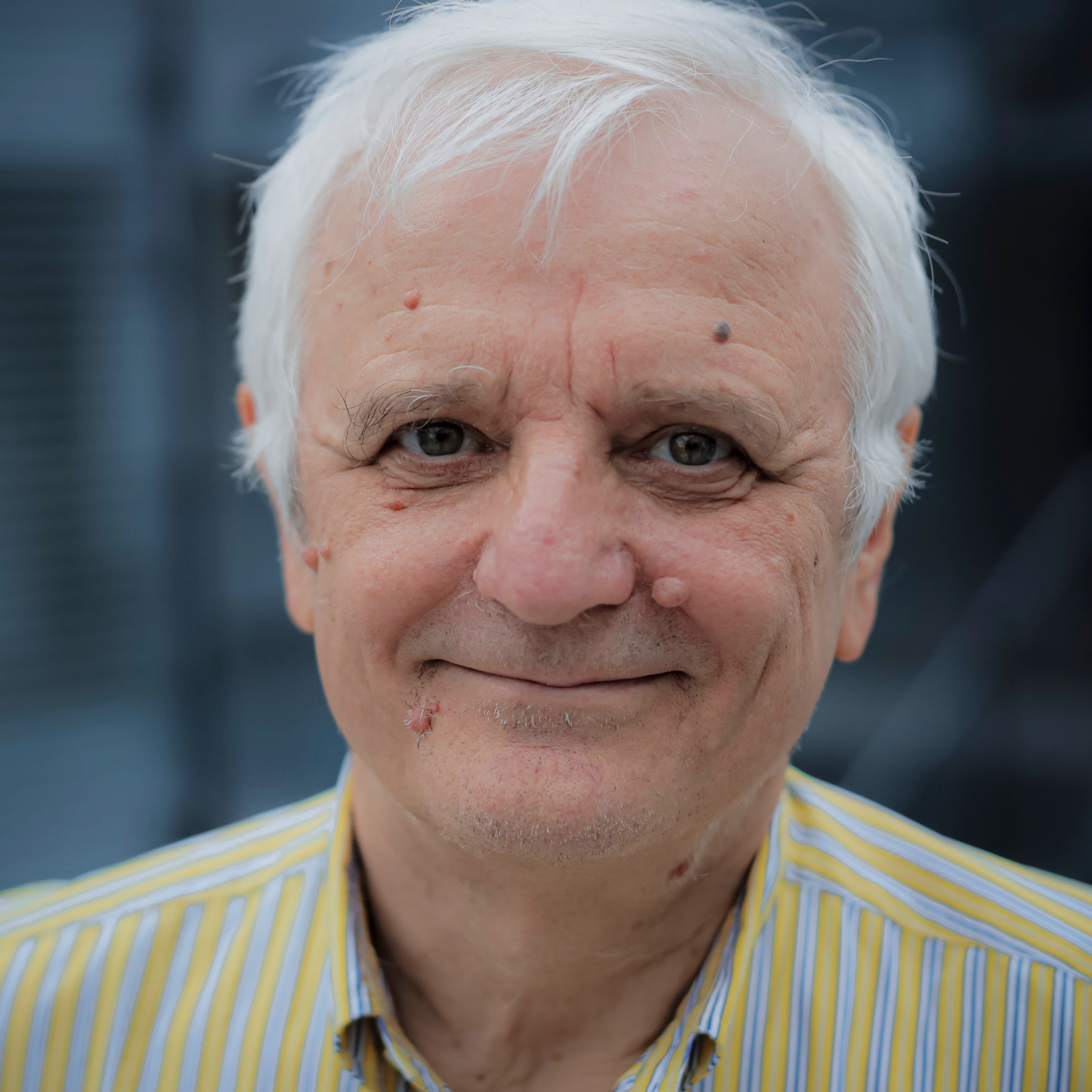}
Boleslaw K. Szymanski is the Claire and Roland Schmitt Distinguished Professor of Computer Science and the Founding Director of the Center for Network Science and Technology, Rensselaer Polytechnic Institute. He was the Director of the Social Cognitive Networks Center in the Network Science Collaborative Technology Alliance, the Principal Investigator in the International Technology Alliance, and in the MilkyWay@home project that models the distribution of dark matter in the Milky Way Galaxy. His projects focus on dynamic processes on networks, groups in social networks, sensor network protocols and algorithms, and large-scale distributed computing. Dr. Szymanski was a visiting professor at Universities of Pennsylvania, Stanford and Wrocław University of Technology. He is a foreign member of Polish Academy of Sciences.
\endbio

\end{document}